\documentclass[11pt,british]{article}
\usepackage[utf8]{inputenc}
\usepackage{geometry}
\usepackage{color}
\usepackage{babel}
\usepackage{booktabs}
\usepackage{mathtools}
\usepackage{amsmath}
\usepackage{amsthm}
\usepackage{amssymb}
\usepackage{cancel}
\usepackage{setspace}
\usepackage{microtype}
\usepackage[unicode=true,
 bookmarks=true,bookmarksnumbered=false,bookmarksopen=false,
 breaklinks=false,pdfborder={0 0 0},pdfborderstyle={},backref=false,colorlinks=true]
 {hyperref}
\hypersetup{pdftitle={Eigenvalues and eigenforms on Calabi–Yau threefolds},
 pdfauthor={Anthony Ashmore},
 colorlinks=true,urlcolor=darkblue,anchorcolor=darkblue,citecolor=darkblue,filecolor=darkblue,linkcolor=darkblue,menucolor=darkblue,linktocpage=true}

\makeatletter

\providecommand{\tabularnewline}{\\}

\usepackage{tikz}
\usetikzlibrary{backgrounds}
\usepackage[framemethod=tikz]{mdframed}
\AtBeginEnvironment{mdframed}{%
\tikzset{every picture/.style={}}%
}
\mdfsetup{roundcorner=.5ex,font=\small}
{\begin{mdframed}[backgroundcolor=white]}%
{\end{mdframed}}
{\begin{mdframed}[backgroundcolor=white]}%
{\end{mdframed}}

\usepackage{slashed}	 	
\usepackage{amsfonts} 		
\usepackage{cite}	
\SetTracking{encoding={*}, shape=sc}{0} 	
\allowdisplaybreaks		
\date{\today} 		
\numberwithin{equation}{section}	

\makeatletter
\g@addto@macro\bfseries{\boldmath}
\makeatother

\usepackage{xcolor}
\definecolor{darkblue}{rgb}{0.1,0.0,0.5}

\makeatletter
\g@addto@macro\@floatboxreset\centering
\makeatother

\let\originalleft\left
\let\originalright\right
\renewcommand{\left}{\mathopen{}\mathclose\bgroup\originalleft}
\renewcommand{\right}{\aftergroup\egroup\originalright}

\usepackage{pgfplots}
\usetikzlibrary{cd}
\usepgfplotslibrary{external}

\tikzexternaldisable

\makeatother

\begin{document}
\global\long\def\rep#1{\boldsymbol{#1}}%
\global\long\def\repb#1{\overline{\boldsymbol{#1}}}%
\global\long\def\dd{\text{d}}%
\global\long\def\ii{\text{i}}%
\global\long\def\ee{\text{e}}%
\global\long\def\Dorf{L}%
\global\long\def\tDorf{\hat{L}}%
\global\long\def\GL#1{\text{GL}(#1)}%
\global\long\def\Orth#1{\text{O}(#1)}%
\global\long\def\SO#1{\text{SO}(#1)}%
\global\long\def\Spin#1{\text{Spin}(#1)}%
\global\long\def\Symp#1{\text{Sp}(#1)}%
\global\long\def\Uni#1{\text{U}(#1)}%
\global\long\def\SU#1{\text{SU}(#1)}%
\global\long\def\Gx#1{\text{G}_{#1}}%
\global\long\def\Fx#1{\text{F}_{#1}}%
\global\long\def\Ex#1{\text{E}_{#1}}%
\global\long\def\ExR#1{\text{E}_{#1}\times\mathbb{R}^{+}}%
\global\long\def\ex#1{\mathfrak{e}_{#1}}%
\global\long\def\gl#1{\mathfrak{gl}_{#1}}%
\global\long\def\SL#1{\text{SL}(#1)}%
\global\long\def\Stab{\operatorname{Stab}}%
\global\long\def\vol{\operatorname{vol}}%
\global\long\def\tr{\operatorname{tr}}%
\global\long\def\ad{\operatorname{ad}}%
\global\long\def\ext{\Lambda}%
\global\long\def\AdS#1{\text{AdS}_{#1}}%
\global\long\def\op#1{\operatorname{#1}}%
\global\long\def\im{\operatorname{im}}%
\global\long\def\re{\operatorname{re}}%
\global\long\def\eqspace{\mathrel{\phantom{{=}}{}}}%
\global\long\def\bZ{\mathbb{Z}}%
\global\long\def\bC{\mathbb{C}}%
\global\long\def\bP{\mathbb{P}}%
\global\long\def\bR{\mathbb{R}}%
\global\long\def\feyn#1{\slashed{#1}}%
\global\long\def\id{\operatorname{id}}%
\global\long\def\ap{\alpha'}%

\begin{titlepage}
\begin{flushright} 
\end{flushright}
\vfill
\begin{center}
{\setstretch{1.3}\Large\bf Eigenvalues and eigenforms on Calabi--Yau threefolds\par} 
\vskip 1cm 
Anthony Ashmore
\vskip 1cm
\textit{\small{}Enrico Fermi Institute \& Kadanoff Center for Theoretical Physics,\\ University of Chicago, Chicago, IL 60637, USA}
\\[.4cm]
\textit{\small{}Sorbonne Universit\'e, CNRS, Laboratoire de Physique Th\'eorique et Hautes Energies,\\ F-75005 Paris, France}
\end{center}
\vfill
\begin{center} \textbf{Abstract} \end{center}
\begin{quote}
We present a numerical algorithm for computing the spectrum of the Laplace--de Rham operator on Calabi--Yau manifolds, extending previous work on the scalar Laplace operator \cite{Braun:2008jp}. Using an approximate Calabi--Yau metric as input, we compute the eigenvalues and eigenforms of the Laplace operator acting on $(p,q)$-forms for the example of the Fermat quintic threefold. We provide a check of our algorithm by computing the spectrum of $(p,q)$-eigenforms on $\mathbb{P}^{3}$.
\end{quote}
\vfill
{\begin{NoHyper}\let\thefootnote\relax\footnotetext{\tt \!\!\!\!\!\!\!\!\!\!\!\! ashmore@uchicago.edu}\end{NoHyper}}
\end{titlepage}

\microtypesetup{protrusion=false} 
\tableofcontents 
\microtypesetup{protrusion=true} 

\section{Introduction}

Despite decades of progress, there are still no examples of realistic string theory compactifications where one can compute masses and couplings fully, and then compare them with experimental data. In the case of Calabi--Yau compactifications~\cite{Candelas:1985en,Strominger:1985it,Strominger:1985ks}, an obvious stumbling block has been the lack of analytic expressions for non-trivial Ricci-flat metrics.\footnote{In a \emph{tour de force} calculation, Kachru et al.~have presented a construction of explicit Ricci-flat metrics on smooth K3 surfaces~\cite{Kachru:2018van,Kachru:2020tat}. Since this relies on a count of BPS states of little string theory, it is not clear to the author if this can be extended to threefolds.} Except in a handful of special examples~\cite{Candelas:1987rx,Candelas:1990rm,Greene:1993vm,Donagi:2006yf,Anderson:2009ge,Braun:2006me}, it is not possible to calculate physical parameters, such as normalised Yukawa couplings or particle masses, without the data of the metric. Non-standard embeddings of the $\Ex 8\times\Ex 8$ heterotic string~\cite{Gross:1984dd,Gross:1985fr,Gross:1985rr} (including heterotic M-theory~\cite{Horava:1995qa,Horava:1996ma,Lukas:1997fg, Lukas:1998yy, Lukas:1998tt}) on Calabi--Yau threefolds have arguably come closest to realistic models of particle physics~(see for example \cite{Braun:2005ux,Bouchard:2005ag,Braun:2005nv,Braun:2005bw,Blumenhagen:2006ux,Blumenhagen:2006wj,Anderson:2009mh,Anderson:2011ns,Anderson:2012yf,Anderson:2013xka,Ovrut:2014rba,Ovrut:2015uea,Deen:2016vyh,Faraggi:1989ka,Cleaver:1998saa,Braun:2011ni,Nibbelink:2015ixa,Nibbelink:2015vha,Dumitru:2018jyb,Dumitru:2018nct,Dumitru:2019cgf,Braun:2017feb,Ovrut:2018qog,Ashmore:2020ocb,Buchbinder:2014qda} and references therein). Unfortunately, these are not ``special'' enough for topological or algebraic arguments alone to determine the physical parameters of the resulting effective theories. It is thus of tremendous importance to better understand explicit Calabi--Yau metrics and their properties.

Recent years have seen a flurry of activity in tackling this problem and related problems numerically. There are now a number of approaches for computing approximate metrics on Kähler manifolds, utilising balanced metrics~\cite{donaldson1,donaldson2}, so-called optimal metrics~\cite{Headrick:2009jz}, position-space methods~\cite{Headrick:2005ch}, and symplectic coordinates~\cite{Doran:2007zn}. These have been used to calculate numerical Calabi--Yau metrics~\cite{Douglas:2006rr,Braun:2007sn}, hermitian Yang--Mills connections~\cite{Douglas:2006hz,Anderson:2011ed,Anderson:2010ke}, Chern--Simons invariants~\cite{Anderson:2020ebu}, curvature expansions~\cite{Cui:2019uhy}, and moduli space metrics~\cite{Keller:2009vj}. Most importantly for us, numerical Calabi--Yau metrics have also been used to compute the spectrum of the \emph{scalar} Laplace operator~\cite{IuliuLazaroiu:2008pk,Braun:2008jp}. The focus of this paper will be extending this to the Laplace operator acting on $(p,q)$-forms.

Our main result is the spectrum of the Laplace operator acting on $(p,q)$-forms for the example of the Fermat quintic threefold. We want to emphasise that it was not necessary to focus on the Fermat quintic -- everything in this paper can be applied to Calabi--Yau metrics on non-simply connected manifolds, quotients or complete intersections, as was done in \cite{Braun:2008jp}. We have implemented our numerical routine in Mathematica~\cite{Mathematica}. All calculations were carried out on an eight-core laptop with 16 GB of RAM. To give an idea of the computation times involved, with three million points for the numerical integration and $k_{\phi}=3$, timings range from roughly half an hour for the $(0,0)$ spectrum to ten hours for the $(1,1)$ spectrum. Thanks to Mathematica's built-in parallelisation for compiled functions, calculations scale well on multi-core hardware.

We begin in Section \ref{sec:The-Laplacian-on} with a review of the Laplace operator and a rough overview of our strategy. We continue in Section \ref{sec:The-spectrum-on-P3} by focusing on $\bP^{3}$, three complex-dimensional projective space. We present the known results for the analytic spectrum of $(p,q)$-eigenforms and compare these with our numerical results, finding agreement and giving us confidence in our algorithm. In Section \ref{sec:The-spectrum-on-quintic}, we apply our numerical method to the Fermat quintic threefold. We find the results for the scalar Laplacian agree with those found previously in \cite{Braun:2008jp}, while the results for general $(p,q)$-forms are new. Note that we present only the eigenvalues in this paper -- we also have access to the approximate eigenforms but we do not find it enlightening to present these explicitly (though this information is important for computing intersection numbers, overlaps of eigenforms, and so on). We discuss our conventions for both real and complex geometry in Appendix \ref{sec:Conventions}, and include some extra plots for the spectrum of the Fermat quintic in Appendix \ref{sec:More-plots}.

\subsubsection*{Future directions}

There are a number of directions for future work. Most importantly, we plan to extend our method to $(p,q)$-forms valued in some vector bundle $V$. When one moves away from the standard embedding, Yukawa couplings and particle masses are computed from integrated products of harmonic $(0,p)$-forms valued in $V$~\cite{Candelas:1985en,Strominger:1985it,Strominger:1985ks}. Together with further progress on moduli stabilisation~\cite{Becker:2005nb,Becker:2006xp,Anderson:2010mh,Melnikov:2011ez,Anderson:2011cza,Anderson:2011ty, Anderson:2014xha,Anderson:2013qca,delaOssa:2014cia,delaOssa:2015maa,Garcia-Fernandez:2015hja,Candelas:2016usb,Ashmore:2018ybe,Blesneag:2018ygh,Ashmore:2019rkx} and non-perturbative superpotentials~\cite{Witten:1999eg,Buchbinder:2002ic,Beasley:2003fx,Basu:2003bq,Braun:2007xh,Braun:2007tp,Braun:2007vy,Bertolini:2014dna,Buchbinder:2016rmw,Buchbinder:2019eal,Buchbinder:2019hyb,Buchbinder:2017azb,Buchbinder:2018hns,Buchbinder:2002pr,Anderson:2015yzz}, finding approximate expressions for these harmonic bundle-valued forms would provide a large step towards our end goal of computing masses and couplings in generic Calabi--Yau compactifications. In another direction, the spectrum of the Laplace operator gives information about both massless and massive modes of the Kaluza--Klein compactification on the corresponding manifold. This gives a large amount of previously unknown ``data'' about both string compactifications and Kähler geometries -- a simple question one might ask is: are there any patterns in this data? At the moment, it is not known how to predict either the eigenvalues or the multiplicities that appear in the spectrum -- perhaps machine-learning techniques used to investigate the string landscape could shed light on this question~\cite{He:2020lbz,He:2020bfv,He:2020fdg,Deen:2020dlf,He:2019vsj,Krefl:2017yox,Ruehle:2017mzq,Carifio:2017bov,Ruehle:2020jrk,Erbin:2020srm,Erbin:2020tks,Halverson:2020opj,Otsuka:2020nsk}. This data might also be useful for analysing the non-BPS properties of world-sheet CFTs with Kähler target spaces. We hope to tackle these questions in the near future.

\section{The Laplacian on \texorpdfstring{$(p,q)$}{(p,q)}-forms\label{sec:The-Laplacian-on}}

Consider a $d$-dimensional compact manifold $X$ without boundary admitting a Riemannian metric $g$. The Laplace--de Rham operator $\Delta$ is given by\footnote{See Appendix \ref{sec:Conventions} for our conventions.}
\begin{equation}
\Delta=\dd\delta+\delta\dd,\label{eq:laplacian}
\end{equation}
where $\delta=\dd^{\dagger}$ is the codifferential or adjoint of $\dd$. We will consider $\Delta$ acting on complex-valued $p$-forms -- since $\Delta$ commutes with complex conjugation, the operator is the same acting on real or complex $p$-forms.

The problem tackled in this paper is to determine the eigenvalues and eigenmodes (eigenforms) of $\Delta$ acting on the space of differential $p$-forms. The eigenforms $\phi$ and eigenvalues $\lambda$ are defined by
\begin{equation}
\Delta\phi=\lambda\phi.\label{eq:eigenforms}
\end{equation}
Eigenforms with eigenvalue zero are known as zero modes or harmonic forms. Recall that with respect to the standard inner product $\langle\,,\,\rangle$ on $p$-forms, $\Delta$ is hermitian and so its eigenvalues are real. Furthermore, taking the inner product and using \eqref{eq:laplacian}, we have
\begin{equation}
\langle\phi,\Delta\phi\rangle=\langle\dd\phi,\dd\phi\rangle+\langle\delta\phi,\delta\phi\rangle.
\end{equation}
Since the inner product is positive semi-definite, the right-hand side is non-negative. Putting these together, the eigenvalues $\lambda$ are real and non-negative. Furthermore, as $X$ is compact, the eigenvalues will take discrete values and the eigenspaces will be finite dimensional. If the metric on $X$ admits any continuous or discrete symmetries, the eigenvalues may be degenerate, so that multiple eigenforms share the same eigenvalue. In what follows, we will denote the $m^{\text{th}}$ eigenvalue by $\lambda_{m}$ and its multiplicity by $\mu_{m}$. Note that the eigenvalues scale with the volume of $X$ measured by $g$ as
\begin{equation}
\lambda\sim\text{Vol}^{-2/d}.
\end{equation}
We always normalise the volume of $X$ to $1$ in the examples that follow.

Let $\{\alpha_{A}\}$ be some basis for the space of complex-valued $p$-forms, where $A$ runs from $1$ to the dimension of the basis, with the usual inner product
\begin{equation}
O_{AB}\equiv\langle\alpha_{A},\alpha_{B}\rangle=\int\star\bar{\alpha}_{A}\wedge\alpha_{B}.\label{eq:inner}
\end{equation}
A complex $p$-form $\beta$ can be expanded in this basis as
\begin{equation}
\beta=\sum_{A}v^{A}\alpha_{A},\label{eq:expansion}
\end{equation}
where we are \emph{not} assuming that the basis is orthonormal with respect to \eqref{eq:inner}. With respect to the basis $\{\alpha_{A}\}$, the matrix elements of the Laplace operator $\Delta_{AB}$ are
\begin{equation}
\Delta_{AB}=\langle\alpha_{A},\Delta\alpha_{B}\rangle=\langle\dd\alpha_{A},\dd\alpha_{B}\rangle+\langle\delta\alpha_{A},\delta\alpha_{B}\rangle,\label{eq:matrix_elements}
\end{equation}
where the explicit form of these terms is given in Appendix \ref{sec:Conventions}. Using the matrix elements \eqref{eq:matrix_elements} and the expansion of the eigenforms as \eqref{eq:expansion}, the eigenvalue equation \eqref{eq:eigenforms} for the eigenforms becomes
\begin{equation}
\sum_{B}\langle\alpha_{A},\Delta\alpha_{B}\rangle v^{B}=\sum_{B}\lambda\langle\alpha_{A},\alpha_{B}\rangle v^{B}.
\end{equation}
This is a ``generalised eigenvalue problem'' of the form $\Delta_{AB}v_{B}=\lambda O_{AB}v_{B}$, where $\Delta_{AB}=\langle\alpha_{A},\Delta\alpha_{B}\rangle$ and $O_{AB}=\langle\alpha_{A},\alpha_{B}\rangle$ measures the lack of orthogonality of the basis $\{\alpha_{A}\}$ with respect to the inner product. The eigenvalues $\lambda$ give the eigenvalues of the Laplace operator \eqref{eq:eigenforms}, while the vector $v^{A}$ describes how the eigenforms can be expanded in the $\{\alpha_{A}\}$ basis. In the remainder of the paper we will mostly focus on the eigenvalues (the spectrum of $\Delta$) -- we also have access to the eigenvectors $v^{A}$ which characterise the eigenforms themselves, but we do not find it enlightening to present these explicitly. This information will be important in future work for computing intersection numbers, overlaps of eigenforms, and so on.

For what follows, we assume $X$ is even-dimensional and admits a complex structure that is compatible with the metric. This ensures that the Laplace operator commutes with the complex structure, so that eigenforms admit a further decomposition into $(p,q)$-forms. The basis $\{\alpha_{A}\}$ should then be thought of as a choice of complex $(p,q)$-forms for fixed values of $p$ and $q$.

As reviewed in \cite{Braun:2008jp}, very little is known about the eigenvalues and eigenfunctions of the scalar Laplacian for a generic metric on a closed manifold. Even less is known about the eigenvalues and eigenforms of the Laplace-de Rham operator (though see \cite{Bonifacio:2020xoc} for recent results on using geometric ``bootstrap'' bounds to constrain overlaps of eigenmodes). As with scalar eigenfunctions, zero modes (with eigenvalue zero) are counted by cohomologies, with the corresponding eigenforms given by the unique harmonic representatives of each class. On real manifolds these are the Betti numbers, while on complex manifolds the number of zero modes is counted by the Hodge numbers. A further observation is that the spectrum of the $p$-form Laplacian is related to both the $(p-1)$ and $(p+1)$ spectrum. Consider a $p$-form $\alpha$ which is an eigenform of the Laplacian with eigenvalue $\lambda>0$. The Hodge decomposition implies that $\alpha$ can be written as $\alpha=\dd\beta+\delta\gamma$ for some $(p-1)$-form $\beta$ and some $(p+1)$-form $\gamma$. Since $\Delta$ commutes with both $\dd$ and $\delta$ and the spaces of $\dd$-exact and $\delta$-exact forms are orthogonal, we can further restrict to $\alpha$ being $\dd$-exact or $\delta$-exact. In the case that $\alpha$ is $\dd$-exact we have
\begin{equation}
\Delta\dd\beta=\dd\Delta\beta\equiv\dd(\lambda\beta),
\end{equation}
which implies that $\beta$ is an $(p-1)$-eigenform of the Laplacian with the same eigenvalue as $\alpha$. The same argument shows that $\gamma$ is a $(p+1)$-eigenform of the Laplacian, also with eigenvalue $\lambda$. Hence, the combined $(p+1)$- and $(p-1)$-form spectrum will contain a single copy of the $\lambda>0$ $p$-form spectrum of $\Delta$. For example, every eigenfunction $\alpha$ with a non-zero eigenvalue leads to a one-form eigenmode $\dd\alpha$ with the same eigenvalue, so that the one-form spectrum contains a copy of the $\lambda>0$ zero-form spectrum. Given the refinement of forms according to complex type and Hodge decompositions for both $\partial$ and $\bar{\partial}$, a similar relation holds for $(p,q)$-forms. Note that if the manifold has a presentation as a symmetric space $G/H$, one can determine the spectrum (the eigenvalues and their multiplicity) using representation theory. We will use this in the next section as a check of our numerical method on $\bP^{3}$. 

A sketch of the algorithm for computing the eigenvalues and eigenforms numerically is as follows:
\begin{enumerate}
\item Choose a complex manifold $X$ together with a hermitian metric that can be computed explicitly. If the chosen metric (such as the Ricci-flat metric on a Calabi--Yau) is not known analytically, a numerical approximation to the metric should be computed.
\item Focusing on the $(p,q)$-form Laplacian, choose a set of forms $\{\alpha_{A}\}$ that spans the space of complex $(p,q)$-forms. The (infinite-dimensional) matrices $\Delta_{AB}$ and $O_{AB}$ can then in principle be computed using some numerical integration scheme over $X$. Obviously, one cannot compute these matrices in practice. Instead one must restrict $\{\alpha_{A}\}$ to a finite set to give an approximate basis. Our choice will be discussed in the next section and is a simple extension of the functions used in \cite{Braun:2008jp,IuliuLazaroiu:2008pk}.
\item Using the approximate basis, which will we also denote by $\{\alpha_{A}\}$, compute the finite-dimensional matrices $\Delta_{AB}$ and $O_{AB}$ and then solve for the eigenvalues $\lambda$ and eigenvectors $v^{A}$.
\end{enumerate}
The result of this algorithm is the approximate eigenvalues and eigenforms of the $(p,q)$-form Laplacian. One can then improve the approximation by increasing the number of points used in the numerical integration step or by increasing the size of the approximate basis $\{\alpha_{A}\}$. As the number of integrations points and the size of the approximating basis tend to infinity, the approximate eigenvalues and eigenforms converge to their exact values.\footnote{These are their exact values with respect to the metric on $X$, which itself might be an approximation.}

\section{The spectrum on \texorpdfstring{$\mathbb{P}^3$}{P3}\label{sec:The-spectrum-on-P3}}

We first apply our method to the complex closed threefold $\bP^{3}$ with the Fubini--Study (FS) metric. The FS metric on $\bP^{3}$ is Kähler and unique up to scale -- we use this freedom to set $\text{Vol}(\bP^{3})=1$ -- and corresponds to the presentation of $\bP^{3}$ as the symmetric space
\begin{equation}
\bP^{3}=\frac{\text{S}^{7}}{\Uni 1}=\frac{\SU 4}{\text{S}(\Uni 3\times\Uni 1)}.
\end{equation}
Thanks to this, the spectra and eigenforms of the Laplacian can be found analytically~\cite{IKEDATANIGUCHI}. This will provide a check of our numerical method and give confidence that the later results for Calabi--Yau threefolds (where analytic results are not available) are correct.

\subsection{Analytic results}

We begin with a review of the analytic results of \cite{IKEDATANIGUCHI} following the presentation in \cite{Braun:2008jp}. We take the FS metric to be $g_{i\bar{j}}=\partial_{i}\bar{\partial}_{\bar{j}}K_{\text{FS}}$, with the Kähler potential given by
\begin{equation}
K_{\text{FS}}=\frac{6^{1/3}}{2\pi}\log\bigl(|z_{0}|^{2}+|z_{1}|^{2}+|z_{2}|^{2}+|z_{3}|^{2}\bigr),
\end{equation}
where $[z_{0}\!:\!z_{1}\!:\!z_{2}\!:\!z_{3}]$ are homogeneous coordinates on $\bP^{3}$. The coefficient of $6^{1/3}$ ensures that the volume of $\bP^{3}$ is normalised to
\begin{equation}
\text{Vol}(\bP^{3})=\int\frac{\omega^{3}}{3!}=1.
\end{equation}

As discussed in \cite{IKEDATANIGUCHI}, the eigenvalues and their multiplicities for the Laplacian on $(p,q)$-forms are characterised by highest weights of $\SU 4$. There are three fundamental weights of $\SU 4$, which we denote by $w_{i}$ with $i=1,2,3$. For ease of notation, we also take $w_{0}=w_{4}=0$. It is then useful to define
\begin{equation}
w(m,r,s)=m(w_{1}+w_{3})+(r-s)w_{1}+w_{s}+w_{3-r+1},
\end{equation}
where the integer parameters $m$, $r$ and $s$ satisfy
\begin{equation}
m+r-s\geq0,\qquad m\geq0.
\end{equation}
Different types of $(p,q)$-forms are characterised by various combinations of highest weights. In particular
\begin{equation}
\begin{aligned}p=q=0: &  &  & w(m,0,0),\\
p=0,\quad0<q<3: &  &  & w(m,0,q)\oplus w(m,0,q+1),\\
0<p<3,\quad q=0: &  &  & w(m,p,0)\oplus w(m,p+1,0),\\
p,q>0,\quad3>p+q: &  &  & w(m,p,q)\oplus w(m,p,q+1)\oplus w(m,p+1,q)\oplus w(m,p+1,q+1),\\
p,q>0,\quad3=p+q: &  &  & w(m,p,q)\oplus w(m,p,q+1)\oplus w(m,p+1,q),\\
p=0,\quad q=3: &  &  & w(m,0,3),\\
p=3,\quad q=0: &  &  & w(m,3,0).
\end{aligned}
\end{equation}
To compute the multiplicities $\mu$, one simply expands $w(k,r,s)$ in $w_{1}$, $w_{2}$ and $w_{3}$, and then computes the dimension of the corresponding $\SU 4$ representation. One finds
\begin{equation}
\mu(m,r,s)=\tfrac{1}{12}(m+1)(m+2)(m+r-s+1)(m+r-s+2)(2m+r-s+3).\label{eq:mu_P3}
\end{equation}
The eigenvalues themselves are given (up to a normalisation factor) by computing the Casimir invariant for the relevant highest weight:\footnote{For example, using the Mathematica package LieART~\cite{1912.10969}, the multiplicity can be computed by taking the coefficients of the $w_{i}$ as $(a,b,c)$ and then using the command \texttt{Dim{[}Irrep{[}A{]}{[}a,b,c{]}{]}}. The eigenvalues themselves can be computed using \texttt{CasimirInvariant{[}Irrep{[}A{]}{[}a,b,c{]}{]}}.} 
\begin{equation}
\lambda(m,r,s)=\frac{4\pi}{6^{1/3}}\left(m^{2}+m(r-s+3)+\tfrac{3}{8}(r-s)(r-s+4)\right),\label{eq:lambda_P3}
\end{equation}
where the coefficient is due to our normalisation of the volume of $\bP^{3}$. 

Note that the above calculation actually gives the eigenspaces of the Laplacian acting on $\ext_{0}^{p,q}$, the space of \emph{primitive} $(p,q)$-forms. The space of $(p,q)$-forms $\ext^{p,q}$ admits a decomposition of the form
\begin{equation}
\ext^{p,q}=\ext_{0}^{p,q}\oplus(\omega\wedge\ext_{0}^{p-1,q-1})\oplus\ldots
\end{equation}
where $\omega$ is the Kähler form. Since wedging or contracting with the Kähler form $\omega$ commutes with $\Delta$, the eigenspaces also respect this decomposition. In practice this means that one has to take into account extra modes when tabulating the eigenvalues and multiplicities. For example, for $(p,q)=(1,1)$, we have the decomposition
\begin{equation}
\ext^{1,1}=\ext_{0}^{1,1}\oplus(\omega\wedge\ext^{0,0}).\label{eq:11_decomp}
\end{equation}
The eigenvalues and multiplicities of $\Delta$ acting on $\ext_{0}^{1,1}$are computed by \eqref{eq:lambda_P3} and \eqref{eq:mu_P3} with
\begin{equation}
(m,r,s)=\bigl\{(m,1,1),(m,1,2),(m,2,1),(m,2,2)\bigr\}.
\end{equation}
We must then combine these with the eigenvalues and multiplicities of $\Delta$ acting on $\ext^{0,0}$, which are computed by taking $(m,r,s)=(m,0,0)$. Note that $\ext_{0}^{1,1}$ on its own does not admit a zero mode -- however we know that there should be a $(1,1)$ zero mode, corresponding to the FS Kähler form $\omega$. This mode comes from taking the zero mode in $\ext^{0,0}$ and wedging with $\omega$, thus giving the required $(1,1)$-form with eigenvalue zero and multiplicity one.

We now want to tabulate the eigenvalues and their multiplicities for the various types of $(p,q)$-forms. Since the Laplacian commutes with complex conjugation and the Hodge star, the eigenvalues (and their multiplicities) obey
\begin{equation}
\lambda^{(p,q)}=\lambda^{(q,p)}=\lambda^{(3-p,3-q)}=\lambda^{(3-q,3-p)}.
\end{equation}
Thanks to this, we only need to compute
\begin{equation}
(p,q)\in\Bigl\{(0,0),(1,0),(2,0),(3,0),(1,1),(2,1)\Bigr\}.
\end{equation}
The eigenvalues and their multiplicities are given in Table \ref{tab:P3_exact}. Note that we denote the exact eigenvalues by $\hat{\lambda}_{m}$ and their multiplicities by $\mu_{m}$. As discussed in Section \ref{sec:The-Laplacian-on}, there is a relation between the massive $p$-form and the $(p-1)$- and $(p+1)$-form spectra. For example, looking at Table \ref{tab:P3_exact}, we see that the $\lambda>0$ $(1,0)$ spectrum is a combination of the $(0,0)$ and $(2,0)$ spectra, with the modes themselves coming from $\partial$ of $(0,0)$ modes and $\partial^{\dagger}$ of $(2,0)$ modes.\footnote{This corresponds to using the Hodge decomposition with respect to $\partial$. The decomposition with respect to $\bar{\partial}$ gives the massive $(1,0)$ as $\bar{\partial}^{\dagger}$ of $(1,1)$ modes.} Similar statements hold for the other $(p,q)$ types.

\begin{table}
\noindent \begin{centering}
\scalebox{0.7}{%
\begin{tabular}{ccccccccccccc}
\toprule 
$(p,q)$ & \multicolumn{2}{c}{$(0,0)$} & \multicolumn{2}{c}{$(1,0)$} & \multicolumn{2}{c}{$(2,0)$} & \multicolumn{2}{c}{$(3,0)$} & \multicolumn{2}{c}{$(1,1)$} & \multicolumn{2}{c}{$(2,1)$}\tabularnewline
\midrule 
$m$ & $\hat{\lambda}_{m}$ & $\mu_{m}$ & $\hat{\lambda}_{m}$ & $\mu_{m}$ & $\hat{\lambda}_{m}$ & $\mu_{m}$ & $\hat{\lambda}_{m}$ & $\mu_{m}$ & $\hat{\lambda}_{m}$ & $\mu_{m}$ & $\hat{\lambda}_{m}$ & $\mu_{m}$\tabularnewline
\midrule
\midrule 
$0$ & $0$ & $1$ & $27.66$ & $15$ & $55.32$ & $45$ & $82.99$ & $35$ & $0$ & $1$ & $27.66$ & $15$\tabularnewline
$1$ & $27.66$ & $15$ & $55.32$ & $45$ & $82.99$ & $35$ & $138.3$ & $189$ & $27.66$ & $30$ & $41.49$ & $20$\tabularnewline
$2$ & $69.16$ & $84$ & $69.16$ & $84$ & $103.7$ & $256$ & $207.5$ & $616$ & $41.49$ & $20$ & $55.32$ & $90$\tabularnewline
$3$ & $124.5$ & $300$ & $103.7$ & $256$ & $138.3$ & $189$ & $290.5$ & $1560$ & $55.32$ & $90$ & $69.16$ & $84$\tabularnewline
$4$ & $193.6$ & $825$ & $124.5$ & $300$ & $166.0$ & $875$ & $387.3$ & $3375$ & $69.16$ & $168$ & $82.99$ & $210$\tabularnewline
\bottomrule
\end{tabular}}
\par\end{centering}
\caption{Exact eigenvalues of $\Delta$ and their multiplicities for $(p,q)$-forms on $\protect\bP^{3}$.\label{tab:P3_exact}}
\end{table}

\subsection{An approximate basis}

As discussed in \cite{Braun:2008jp}, the scalar eigenfunctions of $\Delta$ on $\bP^{3}$ are the $\Uni 1$-invariant spherical harmonics on $\text{S}^{7}$, which can be written as linear combinations of functions of the form
\begin{equation}
\alpha_{A}=\frac{(\text{degree }k_{\phi}\text{ monomial in }z_{i})\overline{(\text{degree }k_{\phi}\text{ monomial in }z_{i})}}{\bigl(|z_{0}|^{2}+|z_{1}|^{2}+|z_{2}|^{2}+|z_{3}|^{2}\bigr)^{k_{\phi}}},\qquad k_{\phi}\geq0,\label{eq:function_basis}
\end{equation}
where the denominator ensures that these functions are well defined on $\bP^{3}$. As we discussed in the previous subsection, it is not necessary for the approximate basis $\{\alpha_{A}\}$ to be orthonormal with respect to the inner product. In fact, it is quicker (and more numerically stable) to use a non-orthonormal basis of ``simple'' functions. For zero-forms, this means taking $\{\alpha_{A}\}$ to be the functions in \eqref{eq:function_basis} at a fixed value of $k_{\phi}$. For $(p,q)$-forms, an appropriate finite basis at degree $k_{\phi}$ can be constructed from similar building blocks.\footnote{As discussed in \cite[Section 5]{Bloch:2010gk}, the following choice comes from the Euler sequence $0\to\Omega^{1}(2)\to\bigoplus_{i=0}^{3}\mathcal{O}(1)\to\mathcal{O}(2)\to0$, which implies that $z_{i}\dd z_{j}-z_{j}\dd z_{i}$ gives a holomorphic section of $\Omega^{1}(2)$. For $(p,0)$-forms at degree $k$, the relevant sequence is
\[
0\to\Omega^{p}(k)\to\bigoplus_{i=0}^{n-1}\mathcal{O}(k-p)\to\Omega^{p-1}(k)\to0,
\]
where $n=\begin{pmatrix}3+1\\
p
\end{pmatrix}$ for $\bP^{3}$ and the map on the right takes the form
\[
\dd z_{i_{1}}\wedge\ldots\wedge\dd z_{i_{p}}\mapsto\sum_{j=1}^{p}(-1)^{j+1}z_{i_{j}}\dd z_{i_{1}}\wedge\ldots\wedge\widehat{\dd z_{i_{j}}}\wedge\ldots\wedge\dd z_{i_{p}},
\]
where the hat indicates omission.} 

Consider the set of $(p,0)$-forms (which are not well-defined by themselves on $\bP^{3}$) of the form
\begin{equation}
\bigl(\text{degree }(k_{\phi}-2p)\text{ monomial}\bigr)\bigwedge^{p}(z_{i}\dd z_{j}-z_{j}\dd z_{i}),
\end{equation}
where $i,j=0,\ldots3$ and $k_{\phi}$ is fixed. From this set, discard any which are meromorphic but not holomorphic, that is, which have $z_{i}$-dependent denominators. The remaining forms can all be written as sums of terms of the form $z_{i}^{k_{\phi}-p}\bigwedge^{p}\dd z_{j}$. From this set, discard any which can be written as linear combinations of the others. We denote the remaining $(p,0)$-forms by $\{\xi_{a}^{(k_{\phi},p)}\}$, where $a$ runs from $1$ to the dimension of the basis. A approximate \emph{finite} basis for $(p,q)$-forms at degree $k_{\phi}$ on $\bP^{3}$ is then given by the set spanned by
\begin{equation}
\frac{\xi_{a}^{(k_{\phi},p)}\wedge\overline{\xi_{b}^{(k_{\phi},q)}}}{\bigl(|z_{0}|^{2}+|z_{1}|^{2}+|z_{2}|^{2}+|z_{3}|^{2}\bigr)^{k_{\phi}}}.
\end{equation}
For example, we have
\begin{equation}
\begin{aligned}\{\xi_{a}^{(k_{\phi},0)}\} & =\text{degree }k_{\phi}\text{ monomials},\\
\{\xi_{a}^{(2,1)}\} & =\{z_{0}\dd z_{1}-z_{1}\dd z_{0},z_{0}\dd z_{2}-z_{2}\dd z_{0},\ldots\},\\
\{\xi_{a}^{(3,1)}\} & =\{z_{0}^{2}\dd z_{1}-z_{0}z_{1}\dd z_{0},z_{0}z_{1}\dd z_{2}-z_{1}z_{2}\dd z_{0},\ldots\},\\
\vdots & =\vdots\\
\{\xi_{a}^{(3,2)}\} & =\{z_{2}\dd z_{0}\wedge\dd z_{1}-z_{1}\dd z_{0}\wedge\dd z_{2}+z_{0}\dd z_{1}\wedge\dd z_{2},\ldots\},
\end{aligned}
\end{equation}
and so on. The dimensions of these sets are
\begin{equation}
\begin{aligned}\dim\{\xi_{a}^{(k_{\phi},0)}\} & =\begin{pmatrix}3+k_{\phi}\\
k_{\phi}
\end{pmatrix}=(1,4,10,20,\ldots)\quad\text{for }k_{\phi}=0,1,\ldots\\
\dim\{\xi_{a}^{(k_{\phi},1)}\} & =(6,20,45,84,\ldots)\quad\text{for }k_{\phi}=2,3,\ldots\\
\dim\{\xi_{a}^{(k_{\phi},2)}\} & =(4,15,36,70\ldots)\quad\text{for }k_{\phi}=3,4,\ldots\\
\dim\{\xi_{a}^{(k_{\phi},3)}\} & =(1,4,10,20\ldots)\quad\text{for }k_{\phi}=4,5,\ldots
\end{aligned}
\end{equation}
The general form of these are
\begin{equation}
\dim\{\xi_{a}^{(k_{\phi},p)}\}=\begin{pmatrix}3+1\\
p
\end{pmatrix}\dim\{\xi_{a}^{(k_{\phi}-p,0)}\}-\dim\{\xi_{a}^{(k_{\phi},p-1)}\},
\end{equation}
where the $3$ indicates we are working on $\bP^{3}$. Note that for $\xi_{a}^{(k_{\phi},p)}$ with $p>0$, one requires $k_{\phi}\geq p+1$. This choice of approximate basis is not orthonormal, and so $\langle\alpha_{A},\alpha_{B}\rangle$ will not be diagonal. We will denote the truncated space of $(p,q)$-forms at degree $k_{\phi}$ by
\begin{equation}
\mathcal{F}_{k_{\phi}}^{p,q}=\left\{ \frac{\xi_{a}^{(k_{\phi},p)}\wedge\overline{\xi_{b}^{(k_{\phi},q)}}}{\bigl(|z_{0}|^{2}+|z_{1}|^{2}+|z_{2}|^{2}+|z_{3}|^{2}\bigr)^{k_{\phi}}}\right\} .\label{eq:approx_basis}
\end{equation}
This choice of basis has the useful property that $\mathcal{F}_{k_{\phi}}^{p,q}\subset\mathcal{F}_{k_{\phi}+1}^{p,q}$, so that one does not lose basis elements when increasing the degree of the basis. The dimension of the approximate basis will be denoted by
\begin{equation}
\dim\mathcal{F}_{k_{\phi}}^{p,q}=\dim\{\xi_{a}^{(k_{\phi},p)}\}\times\dim\{\xi_{a}^{(k_{\phi},q)}\}.
\end{equation}

\subsection{Numerical results}

At this point we have the explicit metric on $\bP^{3}$, which determines the Laplace operator, and an appropriate approximate basis of $(p,q)$-forms. To evaluate the integrals over $X$ that give the matrix elements we need, we have to specify a measure on $X$. This is given simply by
\begin{equation}
\vol=\frac{1}{3!}\omega^{3}.
\end{equation}
Integration over $X$ can then be approximated by a finite sum over $N_{\phi}$ random points
\begin{equation}
\int f\vol\to\frac{1}{N_{\phi}}\sum_{i=1}^{N_{\phi}}f(p_{i}).
\end{equation}
As discussed in \cite{Braun:2008jp}, the integration measure that one implicitly picks when summing over points must agree with $\vol$. This is determined by how one picks the random points. Fortunately, if one picks the random points to be distributed uniformly according to $\SU 4$ on $\bP^{3}$, the resulting measure is the desired FS volume form (since this is the unique $\SU 4$-invariant measure). Note that using a finite number of integration points breaks the $\SU 4$ symmetry, leading to a lifting of the degeneracy in the spectrum -- instead of one eigenvalue $\hat{\lambda}_{m}$ with multiplicity $\mu_{m}$, one finds a cluster of $\mu_{m}$ eigenvalues $\lambda_{m}$ with close but distinct values. In the limit where the number of points goes to infinity and the $\SU 4$ symmetry is restored, the spread in $\lambda_{m}$ will go to zero and they will converge to a single eigenvalue $\hat{\lambda}_{m}$ with multiplicity $\mu_{m}$.

With this all in hand, we can numerically compute the matrix elements $\Delta_{AB}=\langle\alpha_{A},\Delta\alpha_{B}\rangle$ and $O_{AB}=\langle\alpha_{A},\alpha_{B}\rangle$ for various values of $(p,q)$. Following our previous work on numerical metrics~\cite{Ashmore:2019wzb}, we have implemented this in Mathematica~\cite{Mathematica}. As in \cite{Braun:2008jp}, we will vary both $k_{\phi}$, which controls the size of the truncated basis of $(p,q)$-forms, and $N_{\phi}$, the number of points used for the numerical integration. Roughly speaking, as $k_{\phi}$ increases, our finite basis is a better approximation to the honest, infinite-dimensional vector space of $(p,q)$-forms. Since the matrix elements have a finite size, we can only compute as many eigenvalues as the rank of the matrices, which is given by $\dim\mathcal{F}_{k_{\phi}}^{p,q}$. Thus increasing $k_{\phi}$ allows us to compute higher eigenvalues and better approximate the lower ones that already appear in the spectrum for smaller $k_{\phi}$. Increasing the number of points $N_{\phi}$ improves the accuracy of the numerical integration and gets us closer to restoring the $\SU 4$ symmetry of the underlying FS metric. This has the effect of decreasing the spread of the eigenvalues within each cluster that corresponds to a single degenerate eigenvalue in the exact limit, and also improves the accuracy of the eigenvalues higher up the spectrum.

\begin{figure}
\includegraphics{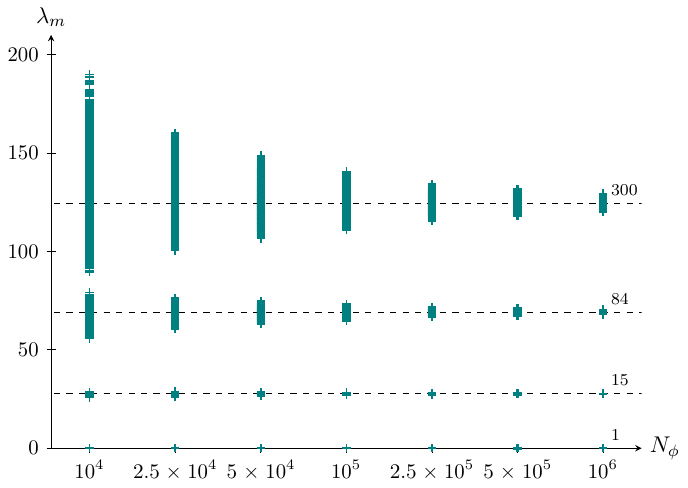}
\caption{A plot of the numerical eigenvalues of the Laplacian on $\protect\bP^{3}$ for $(0,0)$-forms with $k_{\phi}=3$ and varying $N_{\phi}$. The dashed horizontal lines show the values of the exact eigenvalues $\hat{\lambda}_{m}$ from Table \ref{tab:P3_exact}. We also indicate the multiplicity of each eigenvalue, calculated by counting the number of eigenvalues in each grouping, which are in agreement with the exact results.  \label{fig:P3_00_N}}
\end{figure}

\begin{figure}
\includegraphics{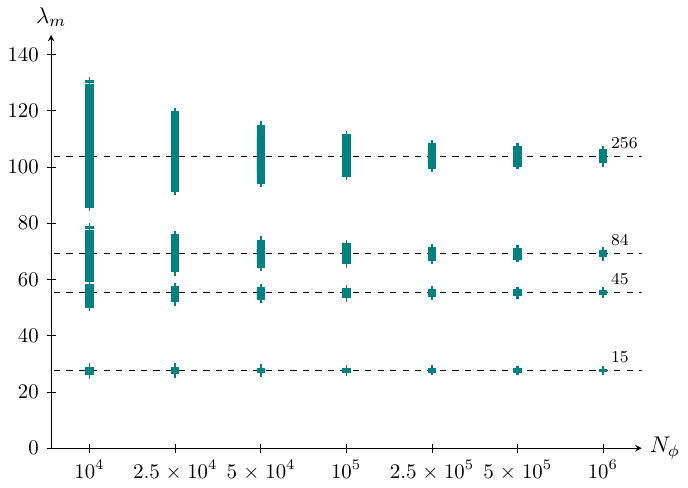}
\caption{A plot of the numerical eigenvalues of the Laplacian on $\protect\bP^{3}$ for $(1,0)$-forms with $k_{\phi}=3$ and varying $N_{\phi}$. The dashed horizontal lines show the values of the exact eigenvalues $\hat{\lambda}_{m}$ from Table \ref{tab:P3_exact}. We also indicate the multiplicity of each eigenvalue, calculated by counting the number of eigenvalues in each grouping, which are in agreement with the exact results. \label{fig:P3_10_N}}
\end{figure}

We show how the $(0,0)$ eigenvalues vary with $N_{\phi}$ for $k_{\phi}=3$ in Figure \ref{fig:P3_00_N}. Since $\dim\mathcal{F}_{3}^{0,0}=400$, we are able to compute the first 400 eigenvalues. We see that as $N_{\phi}$ increases, the approximate eigenvalues tend to their exact values with the spread within each cluster decreasing. Note that these results agree well with a previous plot in \cite[Figure 1]{Braun:2008jp}. We repeat the same calculation for $(1,0)$ eigenvalues with the results in Figure \ref{fig:P3_10_N}. W see that the eigenvalues do indeed converge to their analytic values and that the multiplicities are again correct. Since $\dim\mathcal{F}_{3}^{1,0}=400$, we can compute the first 400 eigenvalues.

\begin{figure}
\includegraphics{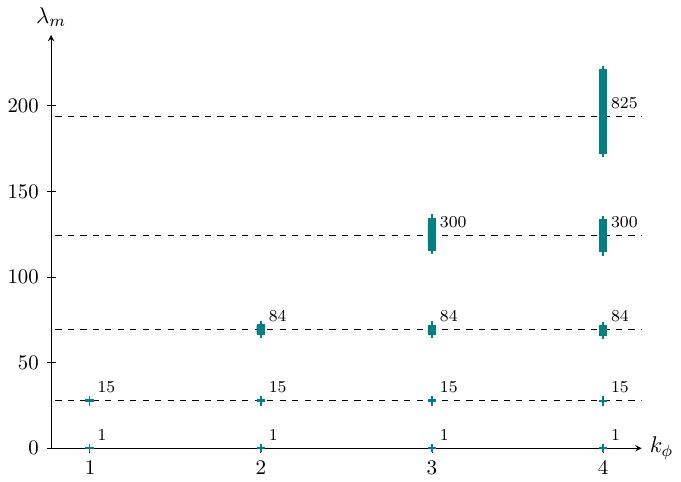}
\caption{A plot of the numerical eigenvalues of the Laplacian on $\protect\bP^{3}$ for $(0,0)$-forms with $N_{\phi}=250{,}000$ and varying $k_{\phi}$. The dashed horizontal lines show the values of the exact eigenvalues $\hat{\lambda}_{m}$ from Table \ref{tab:P3_exact}. We also indicate the multiplicity of each eigenvalue, calculated by counting the number of eigenvalues in each grouping. Notice that the multiplicities do not change as $k_{\phi}$ is increased.  \label{fig:P3_00_k}}
\end{figure}

\begin{figure}
\includegraphics{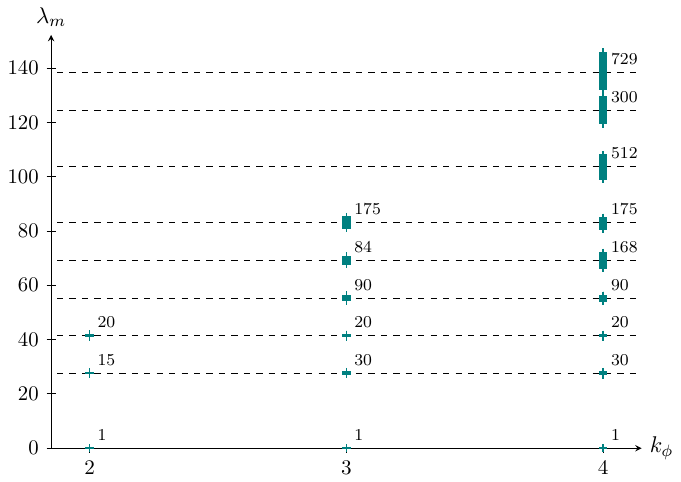}
\caption{A plot of the numerical eigenvalues of the Laplacian on $\protect\bP^{3}$ for $(1,1)$-forms with $N_{\phi}=250{,}000$ and varying $k_{\phi}$. The dashed horizontal lines show the values of the exact eigenvalues $\hat{\lambda}_{m}$ from Table \ref{tab:P3_exact}. We also indicate the multiplicity of each eigenvalue, calculated by counting the number of eigenvalues in each grouping. Notice that the multiplicities can change as $k_{\phi}$ is increased for the reason mentioned in the main text. \label{fig:P3_11_k}}
\end{figure}

We can also vary $k_{\phi}$ while keeping $N_{\phi}$ constant to see how increasing the size of the approximating basis affects the numerical spectrum. We fix $N_{\phi}=250{,}000$ and vary $k_{\phi}$ from $1$ to $4$ for $(p,q)=(0,0)$ and from $2$ to $4$ for $(p,q)=(1,1)$. We display the results in Figure \ref{fig:P3_00_k} and Figure \ref{fig:P3_11_k} respectively. As $k_{\phi}$ increases, higher frequency modes are added to the truncated basis. For the case of $\bP^{3}$ that we are considering, since the $(0,0)$ basis is so closely related to the actual eigenfunctions, increasing $k_{\phi}$ allows us to access higher eigenfunctions with nothing extra appearing lower in the spectrum. For $(1,1)$-eigenforms we have more complicated behaviour, which we comment on below. Note that the lowest-lying $(1,1)$-eigenform has eigenvalue zero (up to numerical error) -- this is the Kähler form for the FS metric on $\bP^{3}$ and it is present for $k_{\phi}\geq2$. The eigenvalue is precisely zero as $\omega$ can written \emph{exactly} as a sum of elements of $\mathcal{F}_{2}^{1,1}$, and since $\mathcal{F}_{2}^{1,1}\subset\mathcal{F}_{k_{\phi}>2}^{1,1}$, the zero mode is present for all $k_{\phi}\geq2$. If one picked a metric different from the FS metric, the Kähler form would be different and would no longer be expressible exactly as a sum of elements of $\mathcal{F}_{2}^{1,1}$. The lowest-lying eigenform would then have eigenvalue close to but not exactly equal to zero. If $k_{\phi}$ were increased, one would find that one could better approximate this new Kähler form and the corresponding eigenvalue would tends to zero. We will see this behaviour again in the case of the Calabi--Yau threefold where the approximate Kähler form cannot be written exactly as a sum of the $(1,1)$ basis forms.

At $k_{\phi}=2$ in the $(1,1)$ spectrum, we see that the first massive mode has multiplicity $\mu_{1}=15$. However, moving to $k_{\phi}=3$ we see that $\mu_{1}=30$, which is the predicted analytic result. How do we understand this? Looking at the $(0,0)$ spectrum in Figure \ref{fig:P3_00_k}, we note that at $k_{\phi}=0$, only the constant zero mode is present, while we pick up the first massive eigenfunction with multiplicity $15$ at $k_{\phi}=1$. Now recall from \eqref{eq:11_decomp} that the space of $(1,1)$-forms splits as a direct sum of primitive $(1,1)$-forms and those that can be written as a function multiplying $\omega$. Since $\omega$ is actually a sum of the $(1,1)$-form basis elements at $k_{\phi}=2$, working at $k_{\phi}=2$ we cannot access $\omega$ wedged with the first $(0,0)$ massive modes. Enlarging our basis to $k_{\phi}=3$, we can then accommodate $\omega$ wedged with the first $(0,0)$ massive mode with multiplicity $15$ (which appears in the $(0,0)$ spectrum for $k_{\phi}\geq1$), which then combines with the $15$ $(1,1)$-eigenforms that were already present at $k_{\phi}=2$. This pattern continues for higher harmonics. For example, at $k_{\phi}=3$ one finds $\mu_{4}=84$, but moving to $k_{\phi}=4$ this changes to $\mu_{4}=168$ due to the contribution of the degree 2 $(0,0)$-mode with $\mu_{2}=84$; this then matches the exact calculation. The general pattern is that one has to compute the $(1,1)$-eigenforms at $k_{\phi}=k+2$ in order to include eigenforms which come from eigenfunctions that appear at degree $k$ in the $(0,0)$ spectrum (effectively because $\omega$ is itself the wedge product of degree 2 basis elements from $\{\xi_{a}^{(2,1)}\}$ and its conjugate). We will see in the case of a Calabi--Yau threefold that the same phenomenon holds -- increasing $k_{\phi}$ can cause the appearance of low-lying eigenforms that were not previously present.

\begin{figure}
\includegraphics{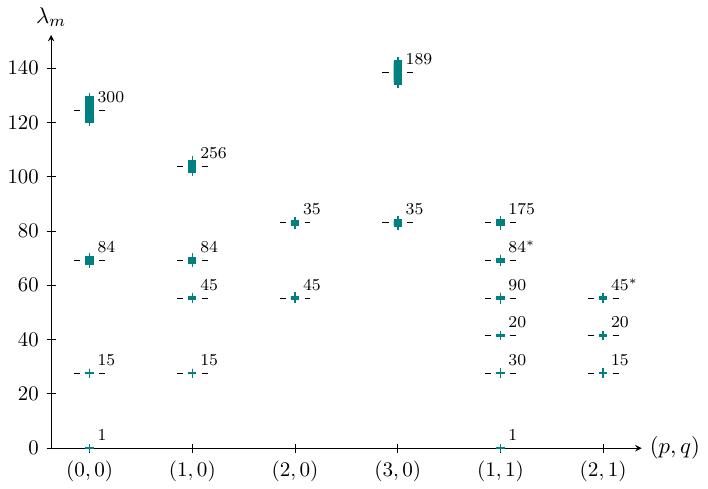}
\caption{A plot of the numerical eigenvalues of the Laplacian on $\protect\bP^{3}$ for $(p,q)$-forms with $N_{\phi}={10}^{6}$ and $k_{\phi}=3$ ($(3,0)$ at $k_{\phi}=5$). The dashed horizontal lines show the values of the exact eigenvalues $\hat{\lambda}_{m}$ from Table \ref{tab:P3_exact}. We also indicate the multiplicity of each eigenvalue, calculated by counting the number of eigenvalues in each grouping. An asterisk on a multiplicity indicates that this does not match the exact value in Table \ref{tab:P3_exact} for the reason discussed in the main text.\label{fig:P3_pq}}
\end{figure}

Rather than plotting the eigenvalues for all values of $(p,q)$ in this way, we show the eigenvalues for the independent values of $(p,q)$ for $N_{\phi}={10}^{6}$ and $k_{\phi}=3$ in Figure \ref{fig:P3_pq}. We also record the approximate values of the eigenvalues and their spread (standard deviation) in Table \ref{tab:P3_approx}, which should be compared with the exact values in Table \ref{tab:P3_exact}. Happily, we see that both the magnitude and multiplicity of our numerically calculated eigenvalues agree with the analytic calculation. Note that we take $k_{\phi}=5$ for $(p,q)=(3,0)$ as this is the minimum degree where one finds at least two clusters of eigenvalues.

\begin{table}
\noindent \begin{centering}
\scalebox{0.7}{%
\begin{tabular}{ccccccccccccc}
\toprule 
$(p,q)$ & \multicolumn{2}{c}{$(0,0)$} & \multicolumn{2}{c}{$(1,0)$} & \multicolumn{2}{c}{$(2,0)$} & \multicolumn{2}{c}{$(3,0)$} & \multicolumn{2}{c}{$(1,1)$} & \multicolumn{2}{c}{$(2,1)$}\tabularnewline
\midrule 
$\dim\mathcal{F}_{k_{\phi}}^{p,q}$ & \multicolumn{2}{c}{$400$} & \multicolumn{2}{c}{$400$} & \multicolumn{2}{c}{$80$} & \multicolumn{2}{c}{$224$} & \multicolumn{2}{c}{$400$} & \multicolumn{2}{c}{$80$}\tabularnewline
\midrule 
$m$ & $\lambda_{m}$ & $\mu_{m}$ & $\lambda_{m}$ & $\mu_{m}$ & $\lambda_{m}$ & $\mu_{m}$ & $\lambda_{m}$ & $\mu_{m}$ & $\lambda_{m}$ & $\mu_{m}$ & $\lambda_{m}$ & $\mu_{m}$\tabularnewline
\midrule
\midrule 
$0$ & $0.000$ & $1$ & $27.66\pm0.10$ & $15$ & $55.32\pm0.28$ & $45$ & $82.95\pm0.58$ & $35$ & $0.000$ & $1$ & $27.66\pm0.11$ & $15$\tabularnewline
$1$ & $27.66\pm0.12$ & $15$ & $55.31\pm0.26$ & $45$ & $82.99\pm0.40$ & $35$ & $138.4\pm2.2$ & $189$ & $27.66\pm0.09$ & $30$ & $41.49\pm0.11$ & $20$\tabularnewline
$2$ & $69.13\pm0.70$ & $84$ & $69.15\pm0.57$ & $84$ &  &  &  &  & $41.49\pm0.09$ & $20$ & $55.33\pm0.18$ & ${45}^{*}$\tabularnewline
$3$ & $124.5\pm2.4$ & $300$ & $103.8\pm1.1$ & $256$ &  &  &  &  & $55.32\pm0.26$ & $90$ &  & \tabularnewline
$4$ &  &  &  &  &  &  &  &  & $69.16\pm0.34$ & ${84}^{*}$ &  & \tabularnewline
$5$ &  &  &  &  &  &  &  &  & $83.00\pm0.50$ & $175$ &  & \tabularnewline
\bottomrule
\end{tabular}}
\par\end{centering}
\noindent \begin{centering}
\par\end{centering}
\caption{Numerical eigenvalues of $\Delta$ and their multiplicity for $(p,q)$-forms on $\protect\bP^{3}$ at $k_{\phi}=3$ ($(3,0)$ at $k_{\phi}=5$) and $N_{\phi}={10}^{6}$. The eigenvalues are the mean of the eigenvalues in a cluster, with the error given by the standard deviation. An asterisk on a multiplicity indicates that this does not match the exact value in Table \ref{tab:P3_exact} for the reason discussed in the main text. \label{tab:P3_approx}}
\end{table}

\section{The spectrum on the Fermat quintic\label{sec:The-spectrum-on-quintic}}

The strategy that we have outlined and applied to $\bP^{3}$ can also be used to investigate the eigenforms of Calabi--Yau metrics. In this note we will focus on the example of the Fermat quintic threefold. We do this in order to compare with previous results for the scalar Laplacian in \cite{Braun:2008jp} -- everything that follows can be easily extended to more complicated examples. As this particular quintic admits a discrete symmetry, we expect the eigenvalues will be degenerate, as we saw with the FS metric on $\bP^{3}$. Note that this symmetry does not simplify the calculation of the eigenvalues (unlike the calculation of the Ricci-flat metric where the symmetry greatly reduces the parameter space over which one minimises). Further details of the Fermat quintic and the numerical integration can be found in \cite[Appendix A]{Ashmore:2019wzb}.

\subsection{A metric on the quintic}

The Fermat quintic $X$ is the hypersurface in $\bP^{4}$ defined by the vanishing locus of the equation
\begin{equation}
Q=z_{0}^{5}+z_{1}^{5}+z_{2}^{5}+z_{3}^{5}+z_{4}^{5},\label{eq:defining}
\end{equation}
where $[z_{0}\!:\!z_{1}\!:\!z_{2}\!:\!z_{3}\!:\!z_{4}]$ are homogeneous coordinates on $\bP^{4}$. The algorithm for computing the eigenvalues and eigenforms on the quintic is essentially the same as we laid out for $\bP^{3}$. The important differences are:
\begin{itemize}
\item The metric on the quintic is not known analytically. Instead it must be computed numerically -- we will use a metric computed using the ``optimal metric'' approach in \cite{Headrick:2009jz}.
\item The approximate basis $\mathcal{F}_{k_{\phi}}^{p,q}$ must take into account that $Q$ vanishes on the hypersurface. When one removes linearly dependent elements to find $\xi_{a}^{(k_{\phi},p)}$, one also removes those that are related by $Q=0$.
\item The random points used for the numerical integration should be chosen according to the exact Calabi--Yau measure (defined by the six-form $\Omega\wedge\bar{\Omega}$) rather than the $\SU 5$-invariant Fubini--Study measure on $\bP^{4}$. Details of this can be found in \cite{Douglas:2006rr,Braun:2007sn}.
\end{itemize}
We compute the metric on the quintic following the method of ``optimal metrics'' in \cite{Headrick:2009jz}. The basic idea is to make an ansatz for the Kähler potential of the Ricci-flat metric, giving a so-called algebraic metric~\cite{donaldson2,tian}, and then to vary the parameters that appear in the ansatz in order to minimise the difference between the approximate metric and the unique Ricci-flat metric. One begins by choosing a basis for the degree-$k_{h}$ polynomials on $\bP^{4}$ in homogeneous coordinates modulo the defining quintic equation. It is simplest to pick the degree-$k_{h}$ monomials, which we denote this basis by $\{s_{\alpha}\}$. This has dimension
\begin{equation}
N(k_{h})=\begin{cases}
\begin{pmatrix}4+k_{h}\\
k_{h}
\end{pmatrix} & 0\leq k_{h}<5,\\
\begin{pmatrix}4+k_{h}\\
k_{h}
\end{pmatrix}-\begin{pmatrix}k_{h}-1\\
k_{h}-5
\end{pmatrix} & k_{h}\geq5,
\end{cases}
\end{equation}
so that $\alpha=1,\ldots,N(k_{h})$. The ansatz for the Kähler potential is then
\begin{equation}
K=\frac{1}{k_{h}\pi}\ln\bigl(s_{\alpha}h^{\alpha\bar{\beta}}\bar{s}_{\bar{\beta}}\bigr),\label{eq:Kahler_potential}
\end{equation}
where $h^{\alpha\bar{\beta}}$ is a positive-definite hermitian matrix of complex numbers that parametrises the metric on $X$. The corresponding metric is given by
\begin{equation}
g_{i\bar{j}}=\partial_{i}\bar{\partial}_{\bar{j}}K,
\end{equation}
which also defines an approximate Kähler form $\omega$. Note that these expressions are written on $\bP^{4}$, but can be simply restricted to $X$. Since $\{s_{\alpha}\}$ is a finite-dimensional approximation to the space of all holomorphic monomials, the metrics given by varying $h^{\alpha\bar{\beta}}$ are a finite subspace of all possible metrics on $X$. The name of the game is then to find $h^{\alpha\bar{\beta}}$ such that $g_{i\bar{j}}$ is the ``best'' metric within the truncated space, where best means closest to the honest Ricci-flat metric.

The key idea of the optimal metrics proposal is how it determines this ``best'' metric. The idea is as follows. One can compute $\det g_{i\bar{j}}$ using the ansatz for $K$, while $\Vert\Omega\Vert^{2}$ can be computed using the exact (and explicitly known~\cite{atiyah1973,Candelas:1987kf}) holomorphic three-form $\Omega$ on the Calabi--Yau. The Kähler metric $g_{i\bar{j}}$ is Ricci-flat if and only if $\det g_{i\bar{j}}$ is proportional to $\Vert\Omega\Vert^{2}$ \emph{pointwise }with the same constant coefficient everywhere on $X$ -- this is simply the Monge--Amp\`ere equation for the metric~\cite{Yau:1978cfy}. The method of \cite{Headrick:2009jz} treats this as a classic minimisation problem where one varies $h^{\alpha\bar{\beta}}$ in order to minimise the error in $\det g_{i\bar{j}}/\Vert\Omega\Vert^{2}=\text{constant}$ on $X$. The optimal metrics found in this way are often orders of magnitude more accurate than those found by Donaldson's iterative method. Since this paper is focused on accurate calculations of the spectrum of the Laplacian, we will use the optimal metrics as our input. These can be computed using the Mathematica package at \cite{headrickmathematicapackage}.

For what follows, we use an approximate metric on the quintic computed at degree ten, that is $k_{h}=10$. For this choice, the $\sigma$-measure of \cite{Douglas:2006rr} that measures how close the approximate metric is to being Ricci-flat is
\begin{equation}
\sigma_{k}\approx7\times{10}^{-5}.
\end{equation}
This has the interpretation that the volume forms defined by $\omega^{3}$ and $\Omega\wedge\bar{\Omega}$ are equal to within 0.007\% (up to an overall normalisation constant).

\subsection{Numerical results}

We present our results for varying $k_{\phi}$ and $N_{\phi}$ for a few cases of $(p,q)$, and then give complete results for $k_{\phi}=3$ and $N_{\phi}=3\times{10}^{6}$. As with $\bP^{3}$, we first have to pick an approximate basis for $(p,q)$-forms. We again take the approximate basis $\mathcal{F}_{k_{\phi}}^{p,q}$ defined in \eqref{eq:approx_basis}. The difference from $\bP^{3}$ is that when we discard the $(p,0)$-forms which can be written as linear combinations of others already in the basis, we also have to take into account that $Q=0$ on the hypersurface. This has the effect of further reducing the size of the basis and is necessary whenever $k_{\phi}-p\geq5$.

\begin{figure}
\includegraphics{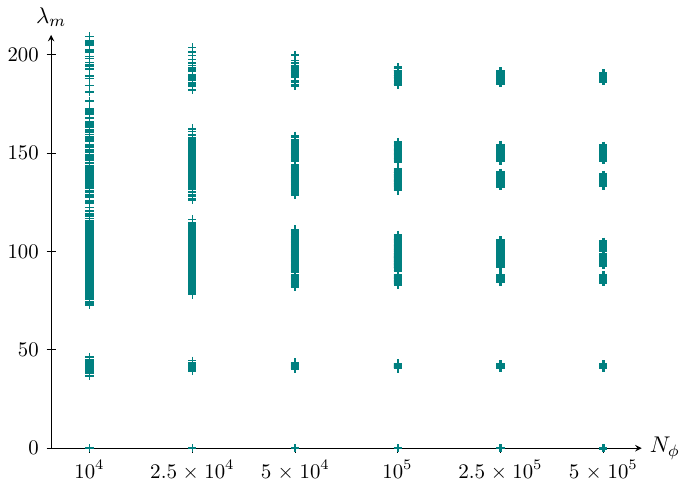}
\caption{A plot of the numerical eigenvalues of the Laplacian on the Fermat quintic for $(0,0)$-forms with $k_{\phi}=2$ and varying $N_{\phi}$. \label{fig:QF_00_N}}
\end{figure}

\begin{figure}
\includegraphics{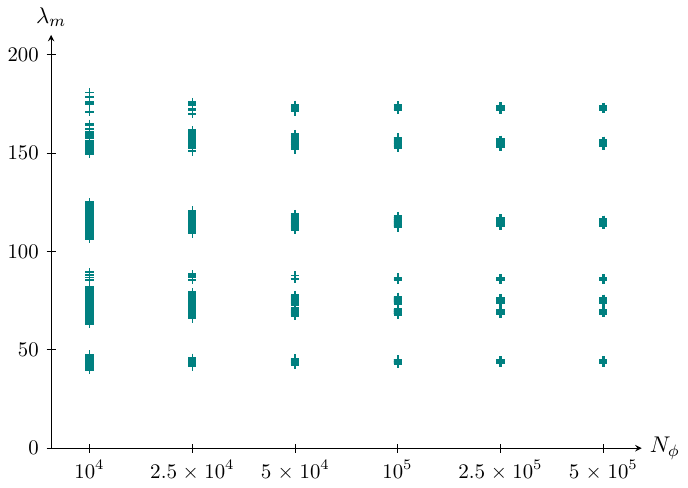}
\caption{A plot of the numerical eigenvalues of the Laplacian on the Fermat quintic for $(1,0)$-forms with $k_{\phi}=2$ and varying $N_{\phi}$. \label{fig:QF_10_N}}
\end{figure}

We show how the $(0,0)$ and $(1,0)$ eigenvalues vary with $N_{\phi}$ for $k_{\phi}=2$ in Figures \ref{fig:QF_00_N} and \ref{fig:QF_10_N}. We see that as $N_{\phi}$ increases, the approximate eigenvalues converge to clusters with smaller and smaller spreads. Since $\dim\mathcal{F}_{2}^{0,0}=225$ we are able to compute the first 225 eigenvalues. Note that the $(0,0)$ results agree well with a previous plot in \cite[Figure 7]{Braun:2008jp}.

We can also vary $k_{\phi}$ while keeping $N_{\phi}$ constant. We fix $N_{\phi}=5\times{10}^{5}$ and vary $k_{\phi}$ from $1$ to $3$ for $(p,q)=(0,0)$ and $2$ to $3$ for $(p,q)=(1,1)$. As $k_{\phi}$ increases, higher-frequency basis forms are added to the truncated basis. Even for the approximate Calabi--Yau metric, there should be a single $(1,1)$ zero mode in the spectrum, corresponding to the Kähler form $\omega$. However, this is not exactly what we see -- instead the lowest-lying mode is a $(1,1)$-form with multiplicity one and a relatively small but non-zero eigenvalue. As $k_{\phi}$ is increased and higher-frequency modes are added to the approximate basis, the lowest-lying $(1,1)$-eigenform has an eigenvalue that gets closer and closer to zero. The reason for this discrepancy is that $\omega$ cannot be written exactly as a sum of elements of the approximate basis -- the metric and thus the (approximate) Kähler form are computed at $k_{h}=10$, and so one imagines needing $k_{\phi}\sim10$ in order to write $\omega$ accurately as a sum of the basis $(1,1)$-forms. Furthermore, looking at the K\"ahler potential in \eqref{eq:Kahler_potential}, the denominator in $\omega$ is of the form $s_{\alpha}h^{\alpha\bar{\beta}}\bar{s}_{\bar{\beta}}=\ee^{\pi k_{h}K}$ whereas elements of $\mathcal{F}_{k_{\phi}}^{1,1}$ have denominator $\bigl(\sum_{i}|z_{i}|^{2}\bigr)^{k_{\phi}}$. If one wanted to better approximate the K\"ahler form, one could use $\ee^{\pi k_{\phi}K}$ as the denominator in \eqref{eq:approx_basis} instead. Since we already have an expression for $\omega$ (from $g_{i\bar{j}}$), we do not find it necessary to implement this change.

\begin{figure}
\includegraphics{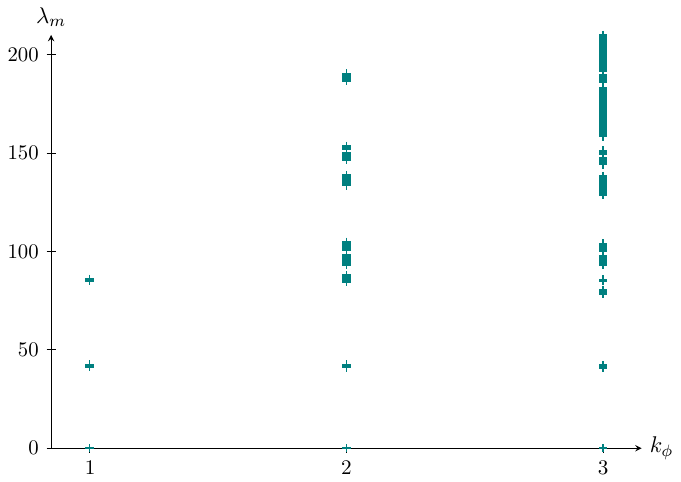}
\caption{A plot of the numerical eigenvalues of the Laplacian on the Fermat quintic for $(0,0)$-forms with $N_{\phi}=5\times{10}^{5}$ and varying $k_{\phi}$. \label{fig:QF_00_k}}
\end{figure}

\begin{figure}
\includegraphics{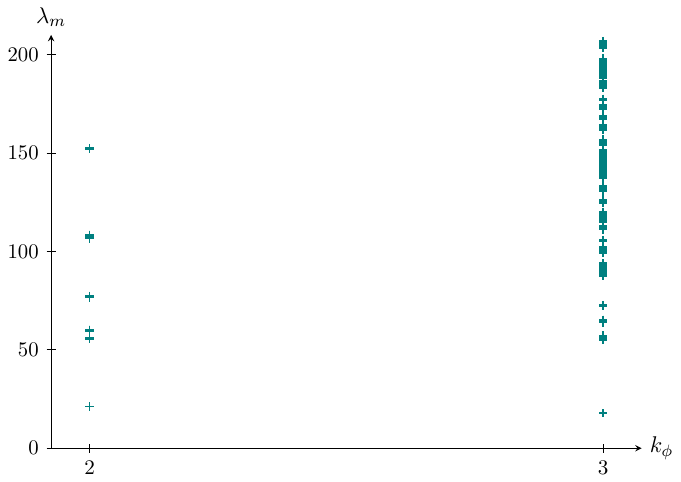}
\caption{A plot of the numerical eigenvalues of the Laplacian on the Fermat quintic for $(1,1)$-forms with $N_{\phi}=5\times{10}^{5}$ and varying $k_{\phi}$. \label{fig:QF_11_k}}
\end{figure}

Finally, we plot the eigenvalues for the independent values of $(p,q)$ for $N_{\phi}=3\times{10}^{6}$ and $k_{\phi}=3$ in Figure \ref{fig:QF_pq}.\footnote{Plots of the spectra for each value of $(p,q)$ can be found in Appendix \ref{sec:More-plots}.} We give the values and the spread of the approximate eigenvalues in Table \ref{tab:QF_approx}. Note that we see there is an isolated $(3,0)$-eigenform with the smallest eigenvalue. This is an approximation to the honest $(3,0)$-form $\Omega$ which should appear as a zero mode. As with the Kähler form $\omega$, if one increases $k_{\phi}$, one finds that the eigenvalue of this isolated mode becomes closer to zero as $\Omega$ is better approximated. Unlike $\omega$, one needs to go to $k_{\phi}\to\infty$ in order to push the numerical eigenvalue down to zero, effectively because we know $\Omega$ is holomorphic, whereas our approximate bases are not. Since we actually have an exact expression for $\Omega$ from a residue theorem, this is not too much of a problem. Unfortunately, for fixed $\dim\mathcal{F}_{k_{\phi}}^{p,q}$ the three-form calculations (both $(3,0)$ and $(2,1)$) are the most time consuming, so we have not pushed to higher values of $k_{\phi}$ or $N_{\phi}$ for these cases on the hardware we have at hand.

\begin{figure}
\includegraphics{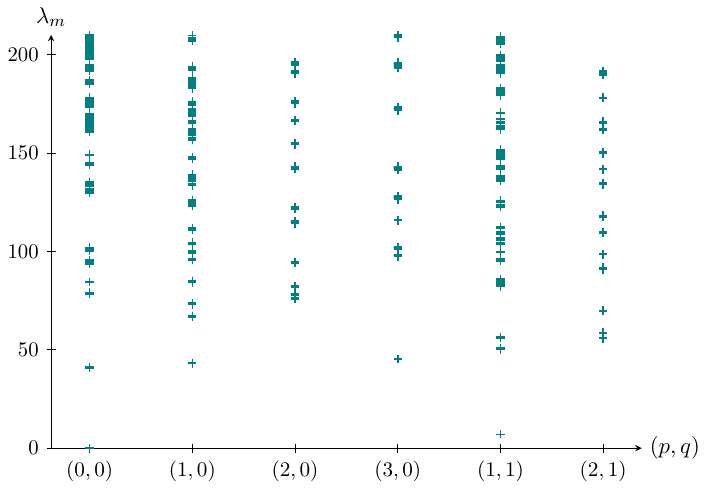}
\caption{A plot of the numerical eigenvalues of the Laplacian on the Fermat quintic for $(p,q)$-forms with $N_{\phi}=3\times{10}^{6}$ and $k_{\phi}=3$ ($(3,0)$ at $k_{\phi}=4$). \label{fig:QF_pq}}
\end{figure}

\begin{table}
\noindent \begin{centering}
\scalebox{0.7}{%
\begin{tabular}{ccccccccccccc}
\toprule 
$(p,q)$ & \multicolumn{2}{c}{$(0,0)$} & \multicolumn{2}{c}{$(1,0)$} & \multicolumn{2}{c}{$(2,0)$} & \multicolumn{2}{c}{$(3,0)$} & \multicolumn{2}{c}{$(1,1)$} & \multicolumn{2}{c}{$(2,1)$}\tabularnewline
\midrule 
$\dim\mathcal{F}_{k_{\phi}}^{p,q}$ & \multicolumn{2}{c}{$1225$} & \multicolumn{2}{c}{$1400$} & \multicolumn{2}{c}{$350$} & \multicolumn{2}{c}{$350$} & \multicolumn{2}{c}{$1600$} & \multicolumn{2}{c}{$400$}\tabularnewline
\midrule 
$m$ & $\lambda_{m}$ & $\mu_{m}$ & $\lambda_{m}$ & $\mu_{m}$ & $\lambda_{m}$ & $\mu_{m}$ & $\lambda_{m}$ & $\mu_{m}$ & $\lambda_{m}$ & $\mu_{m}$ & $\lambda_{m}$ & $\mu_{m}$\tabularnewline
\midrule
\midrule 
$0$ & $0.00$ & $1$ & $43.2\pm0.1$ & $20$ & $76.2\pm0.2$ & $30$ & $45.3$ & $1$ & $7.0$ & $1$ & $56.4\pm0.1$ & $20$\tabularnewline
$1$ & $41.1\pm0.2$ & $20$ & $67.0\pm0.2$ & $30$ & $78.1\pm0.2$ & $30$ & $97.9\pm0.3$ & $20$ & $50.4\pm0.2$ & ${50}^{*}$ & $59.2\pm0.2$ & $20$\tabularnewline
$2$ & $78.7\pm0.3$ & $20$ & $73.3\pm0.2$ & $30$ & $82.0\pm0.1$ & $20$ & $102\pm0.3$ & $20$ & $56.2\pm0.2$ & $20$ & $70.5\pm0.2$ & $30$\tabularnewline
$3$ & $84.5\pm0.1$ & $4$ & $84.6\pm0.2$ & ${34}^{*}$ & $94.5\pm0.2$ & $20$ & $116\pm0.1$ & $4$ & $82.5\pm0.2$ & $60$ & $92.3\pm0.4$ & $60$\tabularnewline
$4$ & $94.6\pm0.6$ & $60$ & $96.0\pm0.2$ & $20$ & $115\pm0.3$ & $40$ & $127\pm0.5$ & $30$ & $84.2\pm0.3$ & $120$ & $99.7\pm0.2$ & $4$\tabularnewline
$5$ & $101\pm1$ & $30$ & $99.6\pm0.3$ & $60$ & $122\pm0.3$ & $30$ & $142\pm0.6$ & $30$ & $85.2\pm0.2$ & $60$ & $111\pm0.4$ & $40$\tabularnewline
\bottomrule
\end{tabular}}
\par\end{centering}
\caption{Numerical eigenvalues of $\Delta$ and their multiplicities for $(p,q)$-forms on the Fermat quintic at $k_{\phi}=3$ ($(3,0)$ at $k_{\phi}=4$) and $N_{\phi}=3\times{10}^{6}$. The eigenvalues are the mean of the eigenvalues in a cluster, with the error given by the standard deviation. Multiplicities with an asterisk denote a cluster of eigenvalues where we do not have sufficient resolution to distinguish between what we suspect are actually two or more clusters (which is implied as the multiplicities are a sum of two of the dimensions appearing in Table \ref{tab:reps}). \label{tab:QF_approx}}
\end{table}

As discussed in \cite{Braun:2008jp}, the zero-locus of the defining equation \eqref{eq:defining} is left invariant by a large discrete symmetry, causing degeneracy of the eigenvalues of the scalar Laplacian. The same is also true for the Laplace--de Rham operator on $(p,q)$-forms. The full non-abelian symmetry that acts linearly on the basis of $(p,q)$-forms (and hence the eigenforms) is
\begin{equation}
\overline{\text{Aut}}(Q)=\bZ_{2}\ltimes\text{Aut}(Q)=(S_{5}\times\bZ_{2})\ltimes(\bZ_{5})^{4},
\end{equation}
which can be thought of, roughly, as symmetric permutations of the $z_{i}$, complex conjugation, and various phase rotations of the $z_{i}$ by fifth roots of unity. There are a finite number of irreducible representations of this discrete group -- we give these in Table \ref{tab:reps}.\footnote{The author thanks F.~Ruehle for explaining how to compute these representations while collaborating on a related project~\cite{2103.07472}. Note that these numbers disagree slightly with \cite[Table 3]{Braun:2008jp}.} Looking at Table \ref{tab:QF_approx}, we see that the multiplicities of the approximate eigenvalues all appear in Table \ref{tab:reps}. That is, the eigenspaces of the $(p,q)$-form Laplacian do indeed live in irreducible representations of $\overline{\text{Aut}}(Q)$, as was the case for the scalar Laplacian in \cite{Braun:2008jp}. This is an independent check that our numerical algorithm is correct.
\begin{table}
\noindent \begin{centering}
\begin{tabular}{c|ccccccccccccc}
Dimension & $1$ & $4$ & $5$ & $6$ & $10$ & $20$ & $24$ & $30$ & $40$ & $48$ & $60$ & $80$ & $120$\tabularnewline
\hline 
\# of irreps & $4$ & $4$ & $4$ & $2$ & $4$ & $18$ & $2$ & $12$ & $12$ & $2$ & $10$ & $2$ & $4$\tabularnewline
\end{tabular}
\par\end{centering}
\caption{The distinct dimensions of the irreducible representations of $\overline{\text{Aut}}(Q)$, together with the number of irreps that have a given dimension.\label{tab:reps}}
\end{table}

\begin{figure}
\includegraphics{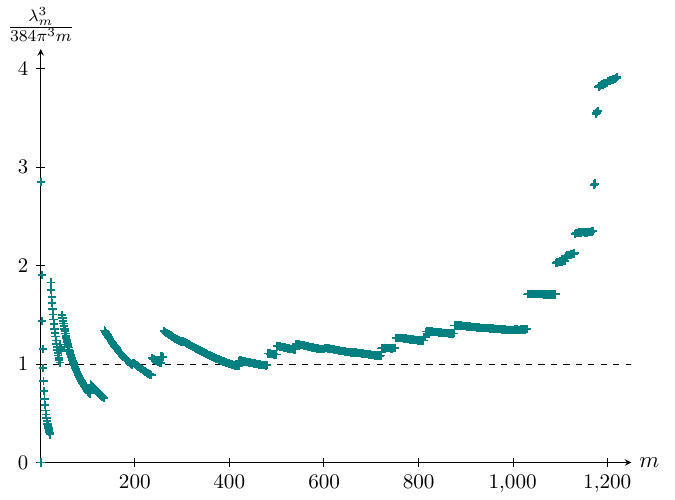}
\caption{A check of Weyl's law for the numerical eigenvalues of the Laplacian on the Fermat quintic for $(0,0)$-forms with $k_{\phi}=3$ and $N_{\phi}=3\times{10}^{6}$. \label{fig:QF_00_weyl}}
\end{figure}

We can also check that our volume normalisation of $\op{Vol}(X)=1$ has been implemented correctly by examining Weyl's law. On a Riemannian manifold $X$ of real dimension $d$, the eigenvalues should grow as
\begin{equation}
\lim_{m\to\infty}\frac{\lambda_{m}^{3}}{m}=\frac{(4\pi)^{d/2}\Gamma(1+d/2)}{\op{Vol}(X)}.
\end{equation}
For the Fermat quintic with unit volume, this means one should find
\begin{equation}
\lim_{m\to\infty}\frac{\lambda_{m}^{3}}{m}=384\pi^{3}.
\end{equation}
Note that degenerate eigenvalues are included in this series according to their multiplicities. Since our numerical spectrum is not exactly degenerate, we simply use the eigenvalues as they are. We give a check of this in Figure \ref{fig:QF_00_weyl}, where we have plotted $\lambda_{m}^{3}/(384\pi^{3}m)$ against $m$ for the $(0,0)$ eigenvalues. Weyl's law requires that this asymptotes to $1$ as $m\to\infty$. Notice that this behaviour is indeed present, implying our volume normalisation is correct. However, around $m\sim700$, the eigenvalues begin to grow at a rate faster than that suggested by Weyl's law. This is a sign that eigenfunctions higher up the spectrum are less well approximated by the finite basis of functions at $k_{\phi}=3$. One can see this behaviour in more detail by varying $k_{\phi}$, as was done in \cite{Braun:2008jp}.

\subsection*{Acknowledgements}

I thank Burt Ovrut, Clay Córdova, Matthew Headrick and Fabian Ruehle for interesting discussions on related work, and Yang-Hui He for previous collaboration on numerical metrics. This work was completed while I was supported by the European Union's Horizon 2020 research and innovation programme under the Marie Sk\l{}odowska-Curie grant agreement No.~838776. I also acknowledge support from NSF Grant No.~PHY2014195 and the Kadanoff Center for Theoretical Physics.

\appendix

\section{Conventions\label{sec:Conventions}}

We denote real coordinate indices by $\{a,b,\ldots\}$ and complex coordinates by $\{i,j,\ldots\}$ and $\{\bar{i},\bar{j},\ldots\}$. Our conventions are as follows.

We define the usual Levi-Civita connection by
\begin{equation}
\Gamma_{bc}^{a}=\tfrac{1}{2}g^{ad}(\partial_{b}g_{dc}+\partial_{c}g_{bd}-\partial_{d}g_{bc}),
\end{equation}
with the covariant derivative on vectors and one-forms
\begin{equation}
\nabla_{a}v^{b}=\partial_{a}v^{b}+\Gamma_{ac}^{b}v^{c},\qquad\nabla_{a}\alpha_{b}=\partial_{a}\alpha_{b}-\Gamma_{ab}^{c}\alpha_{c}.
\end{equation}
The Riemann tensor is defined by
\begin{equation}
\begin{aligned}[][\nabla_{a},\nabla_{b}]v_{c} & =R_{abc}{}^{d}v_{d},\\
R_{abc}{}^{d} & =-2\partial_{[a}\Gamma_{b]c}^{d}+2\Gamma_{c[a}^{e}\Gamma_{b]e}^{d},
\end{aligned}
\end{equation}
with symmetries
\begin{equation}
R_{abcd}=R_{cdab}=-R_{bacd}=-R_{abdc},\qquad R_{[abc]}{}^{d}=0,\qquad\nabla_{[a}R_{bc]d}{}^{e}=0.
\end{equation}
The Ricci tensor and scalar are defined as
\begin{equation}
R_{ab}=R_{abc}{}^{b},\qquad R=R_{a}{}^{a}.
\end{equation}
On a $d$-dimensional Euclidean manifold, the Hodge star and wedge product for a $p$-form $\alpha$ and $q$-form $\beta$ are given by
\begin{align}
(\star\alpha)_{a_{1}\ldots a_{d-p}} & =\frac{1}{p!}\epsilon_{a_{1}\ldots a_{d-p}}{}^{b_{1}\ldots b_{p}}\alpha_{b_{1}\ldots b_{p}},\\
\star^{2}\alpha & =(-1)^{p(d-p)}\alpha,\\
(\alpha\wedge\beta)_{a_{1}\ldots a_{p}b_{1}\ldots b_{q}} & =\frac{(p+q)!}{p!q!}\alpha_{[a_{1}\ldots a_{p}}\beta_{b_{1}\ldots b_{q}]},
\end{align}
where
\begin{equation}
\epsilon_{1\ldots d}=\sqrt{\det g_{ab}}=g^{1/2},\qquad\epsilon^{1\ldots d}=g^{-1/2}.
\end{equation}
The exterior derivative and codifferential are
\begin{align}
\begin{split}\dd v & =\frac{1}{(p+1)!}(\dd v)_{a_{1}\ldots a_{p+1}}\dd x^{a_{1}}\wedge\ldots\wedge\dd x^{a_{p+1}}\\
 & =\frac{1}{(p+1)!}(p+1)\partial_{[a_{1}}v_{a_{2}\ldots a_{p+1}]}\dd x^{a_{1}}\wedge\ldots\wedge\dd x^{a_{p+1}}
\end{split}
\\
\begin{split}\delta v & =(-1)^{dp}\star d\star v\\
 & =\frac{1}{(p-1)!}(\delta v)_{a_{2}\ldots a_{p}}\dd x^{a_{2}}\wedge\ldots\wedge\dd x^{a_{p}}\\
 & =-\frac{1}{(p-1)!}\nabla_{a_{1}}v^{a_{1}}{}_{a_{2}\ldots a_{p}}\dd x^{a_{2}}\wedge\ldots\wedge\dd x^{a_{p}}.
\end{split}
\end{align}
The natural inner product on $p$-forms is taken to be
\begin{equation}
\langle v,w\rangle=\int\star\bar{v}\wedge w=\frac{1}{p!}\int\dd^{d}x\sqrt{g}\,g^{a_{1}b_{1}}\ldots g^{a_{p}b_{p}}\bar{v}_{a_{1}\ldots a_{p}}w_{b_{1}\ldots b_{p}}.\label{eq:inner-1}
\end{equation}
The Laplace--de Rham operator is then simply
\begin{equation}
\Delta=\dd\delta+\delta\dd.
\end{equation}

As discussed in the main text, there are two matrices that we are interested in computing, namely $\langle\alpha_{A},\Delta\alpha_{B}\rangle$ and $\langle\alpha_{A},\alpha_{B}\rangle$ for $\{\alpha_{A}\}$ a finite basis of $(p,q)$-forms. The measure of non-orthogonality of the basis, $\langle\alpha_{A},\alpha_{B}\rangle$, can be computed straightforwardly using \eqref{eq:inner-1}. The matrix elements of $\Delta$ are most easily computed using
\begin{equation}
\langle\alpha_{A},\Delta\alpha_{B}\rangle=\langle\dd\alpha_{A},\dd\alpha_{B}\rangle+\langle\delta\alpha_{A},\delta\alpha_{B}\rangle.
\end{equation}

We further assume that the manifold of interest is Kähler, such as $\bP^{n}$ or a Calabi--Yau. Recall that a Kähler metric satisfies
\begin{equation}
g_{i\bar{j}}=g_{\bar{j}i},\qquad g_{ij}=g_{\bar{i}\bar{j}}=0,\qquad\partial_{k}g_{i\bar{j}}=\partial_{i}g_{k\bar{j}}.
\end{equation}
where the first and last of these identities follows from $g_{i\bar{j}}=\partial_{i}\bar{\partial}_{\bar{j}}K$. The connection symbols $\Gamma_{bc}^{a}$ have only pure holomorphic or antiholomorphic components, namely
\begin{equation}
\Gamma_{jk}^{i}=g^{i\bar{l}}\partial_{j}g_{k\bar{l}},\qquad\Gamma_{\bar{j}\bar{k}}^{\bar{i}}=g^{\bar{i}l}\bar{\partial}_{\bar{j}}g_{\bar{k}l}\equiv\overline{\Gamma_{jk}^{i}},
\end{equation}
with the non-pure connection symbols vanishing. Covariant derivatives are then given by
\begin{equation}
\begin{aligned}\nabla_{i}v^{j} & =\partial_{i}v^{j}+\Gamma_{ik}^{j}v^{k}, & \nabla_{\bar{i}}v^{\bar{j}} & =\bar{\partial}_{\bar{i}}v^{\bar{j}}+\Gamma_{\bar{i}\bar{k}}^{\bar{j}}v^{\bar{k}},\\
\nabla_{i}v^{\bar{j}} & =\partial_{i}v^{\bar{j}}, & \nabla_{\bar{i}}v^{j} & =\bar{\partial}_{\bar{i}}v^{j}.
\end{aligned}
\end{equation}
The non-vanishing Riemann tensor components are
\begin{equation}
R_{i\bar{j}\bar{k}}{}^{\bar{l}}=-R_{\bar{j}i\bar{k}}{}^{\bar{l}}=-\partial_{i}\Gamma_{\bar{j}\bar{k}}^{\bar{l}},\qquad R_{\bar{i}jk}{}^{l}=-R_{j\bar{i}k}{}^{l}=-\bar{\partial}_{\bar{i}}\Gamma_{jk}^{l}\equiv\overline{R_{i\bar{j}\bar{k}}{}^{\bar{l}}}.\label{eq:Riemann-2}
\end{equation}
with the Ricci tensor given by
\begin{equation}
R_{i\bar{j}}=-\partial_{i}\Gamma_{\bar{k}\bar{j}}^{\bar{k}}.
\end{equation}
Note that in complex coordinates the determinants of the real and hermitian metric are related by
\begin{equation}
\sqrt{g}=\sqrt{\det g_{ab}}=2^{d/2}\det g_{i\bar{j}}.
\end{equation}

Let us see explicitly how to compute the matrix elements $\langle\alpha_{A},\Delta\alpha_{B}\rangle$ and $\langle\alpha_{A},\alpha_{B}\rangle$ for a few examples. For the case of $(0,0)$-forms or simply functions, the inner product is
\begin{align}
\langle\alpha_{A},\alpha_{B}\rangle & =\int\dd^{d}x\sqrt{g}\,\bar{\alpha}_{A}\alpha_{B}.
\end{align}
 The matrix elements of $\Delta$ can be calculated by
\begin{equation}
\begin{aligned}\langle\alpha_{A},\Delta\alpha_{B}\rangle & =\langle\dd\alpha_{A},\dd\alpha_{B}\rangle\\
 & =\int\dd^{d}x\sqrt{g}\,g^{ab}(\dd\bar{\alpha}_{A})_{a}(\dd\alpha_{B})_{b}\\
 & =\int\dd^{d}x\sqrt{g}\,\left(g^{i\bar{j}}(\dd\bar{\alpha}_{A})_{i}(\dd\alpha_{B})_{\bar{j}}+g^{\bar{j}i}(\dd\bar{\alpha}_{A})_{\bar{j}}(\dd\alpha_{B})_{i}\right)\\
 & =\int\dd^{d}x\sqrt{g}\,g^{i\bar{j}}\left(\partial_{i}\bar{\alpha}_{A}\bar{\partial}_{\bar{j}}\alpha_{B}+\bar{\partial}_{\bar{j}}\bar{\alpha}_{A}\partial_{i}\alpha_{B}\right).
\end{aligned}
\end{equation}
Note however that there is a simplification thanks to the Kähler structure on $X$ -- the de Rham Laplacian and the Dolbeault Laplacians are proportional:
\begin{equation}
\Delta=2\Delta_{\partial}=2\Delta_{\bar{\partial}}.
\end{equation}
Since most of the $(p,q)$ cases that we consider have $q=0$ and $\bar{\partial}^{\dagger}$ annihilates $(p,0)$-forms, it greatly simplifies calculations to use the $\bar{\partial}$-Laplacian. On functions we then have
\begin{equation}
\begin{aligned}\langle\alpha_{A},\Delta\alpha_{B}\rangle & =2\langle\alpha_{A},\Delta_{\bar{\partial}}\alpha_{B}\rangle\\
 & =2\langle\bar{\partial}\alpha_{A},\bar{\partial}\alpha_{B}\rangle+2\cancel{\langle\bar{\partial}^{\dagger}\alpha_{A},\bar{\partial}^{\dagger}\alpha_{B}\rangle}\\
 & =2\int\dd^{d}x\sqrt{g}\,g^{i\bar{j}}(\partial\bar{\alpha}_{A})_{i}(\bar{\partial}\alpha_{B})_{\bar{j}}\\
 & =2\int\dd^{d}x\sqrt{g}\,g^{i\bar{j}}\partial_{i}\bar{\alpha}_{A}\bar{\partial}_{\bar{j}}\alpha_{B}.
\end{aligned}
\end{equation}
For $(1,0)$-forms
\begin{equation}
\begin{aligned}\langle\alpha_{A},\Delta\alpha_{B}\rangle & =2\langle\bar{\partial}\alpha_{A},\bar{\partial}\alpha_{B}\rangle+2\cancel{\langle\bar{\partial}^{\dagger}\alpha_{A},\bar{\partial}^{\dagger}\alpha_{B}\rangle}\\
 & =\int\dd^{d}x\sqrt{g}\,g^{a_{1}b_{1}}g^{a_{2}b_{2}}(\partial\bar{\alpha}_{A})_{a_{1}a_{2}}(\bar{\partial}\alpha_{B})_{b_{1}b_{2}}\\
 & =2\int\dd^{d}x\sqrt{g}\,g^{i_{1}\bar{j}_{1}}g^{j_{2}\bar{i}_{2}}\partial_{i_{1}}\bar{\alpha}_{A,\bar{i}_{2}}\bar{\partial}_{\bar{j}_{1}}\alpha_{B,j_{2}}
\end{aligned}
\end{equation}
For $(1,1)$-forms, we have
\begin{equation}
\langle\alpha_{A},\Delta\alpha_{B}\rangle=2\langle\bar{\partial}\alpha_{A},\bar{\partial}\alpha_{B}\rangle+2\langle\bar{\partial}^{\dagger}\alpha_{A},\bar{\partial}^{\dagger}\alpha_{B}\rangle
\end{equation}
where
\begin{align}
\begin{split}\langle\bar{\partial}\alpha_{A},\bar{\partial}\alpha_{B}\rangle & =\frac{1}{3!}\int\dd^{d}x\sqrt{g}\,g^{a_{1}b_{1}}g^{a_{2}b_{2}}g^{a_{3}b_{3}}(\partial\bar{\alpha}_{A})_{a_{1}a_{2}a_{3}}(\bar{\partial}\alpha_{B})_{b_{1}b_{2}b_{3}}\\
 & =\frac{1}{2}\int\dd^{d}x\sqrt{g}\,g^{i_{1}\bar{j}_{1}}g^{\bar{i}_{2}j_{2}}g^{i_{3}\bar{j}_{3}}(\partial\bar{\alpha}_{A})_{i_{1}\bar{i}_{2}i_{3}}(\bar{\partial}\alpha_{B})_{\bar{j}_{1}j_{2}\bar{j}_{3}},
\end{split}
\\
\begin{split}\langle\bar{\partial}^{\dagger}\alpha_{A},\bar{\partial}^{\dagger}\alpha_{B}\rangle & =\int\dd^{d}x\sqrt{g}\,g^{a_{1}b_{1}}(\partial^{\dagger}\bar{\alpha}_{A})_{a_{1}}(\bar{\partial}^{\dagger}\alpha_{B})_{b_{1}}\\
 & =\int\dd^{d}x\sqrt{g}\,g^{\bar{i}_{1}j_{1}}(\partial^{\dagger}\bar{\alpha}_{A})_{\bar{i}_{1}}(\bar{\partial}^{\dagger}\alpha_{B})_{j_{1}}.
\end{split}
\end{align}
The components we need are given by
\begin{equation}
(\bar{\partial}\alpha)_{\bar{j}_{1}j_{2}\bar{j}_{3}}=\bar{\partial}_{\bar{j}_{1}}\alpha_{j_{2}\bar{j}_{3}}-\bar{\partial}_{\bar{j}_{3}}\alpha_{j_{2}\bar{j}_{1}},\qquad(\bar{\partial}^{\dagger}\alpha)_{j_{1}}=g^{k_{1}\bar{i}_{1}}\nabla_{k_{1}}\alpha{}_{j_{1}\bar{i}_{1}},
\end{equation}
One can derive expressions for the remaining cases in a similar fashion.

\section{More plots\label{sec:More-plots}}

\begin{figure}
\includegraphics{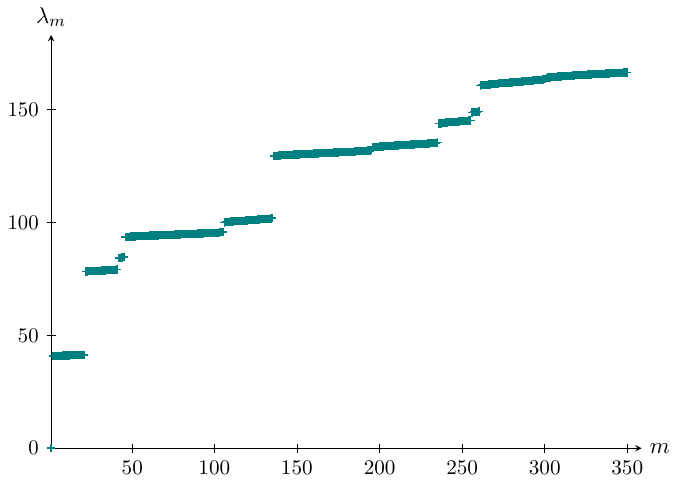}
\caption{A plot of the numerical eigenvalues of the Laplacian on the Fermat quintic for $(0,0)$-forms with $k_{\phi}=3$ and $N_{\phi}=3\times{10}^{6}$. \label{fig:extra1}}
\end{figure}

In this appendix we present more plots of the eigenvalues for the $(p,q)$ Laplacian on the Fermat quintic. In particular, in Figures \ref{fig:extra1} to \ref{fig:extra2} we give the eigenvalues $\lambda_{m}$ plotted against $m$, the eigenvalue number. In all cases, one can clearly see the clusters of nearly degenerate eigenvalues.

\begin{figure}
\includegraphics{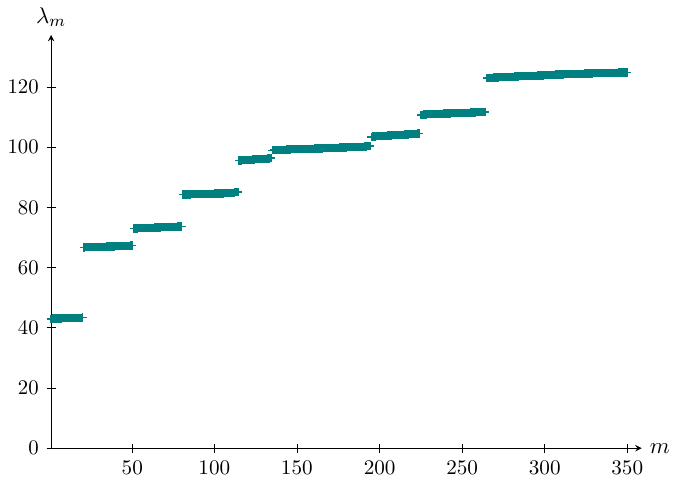}
\caption{A plot of the numerical eigenvalues of the Laplacian on the Fermat quintic for $(1,0)$-forms with $k_{\phi}=3$ and $N_{\phi}=3\times{10}^{6}$.}
\end{figure}

\begin{figure}
\includegraphics{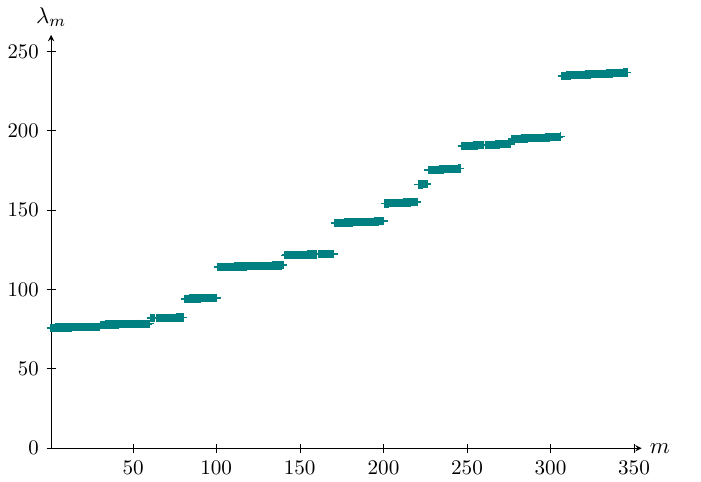}
\caption{A plot of the numerical eigenvalues of the Laplacian on the Fermat quintic for $(2,0)$-forms with $k_{\phi}=3$ and $N_{\phi}=3\times{10}^{6}$.}
\end{figure}

\begin{figure}
\includegraphics{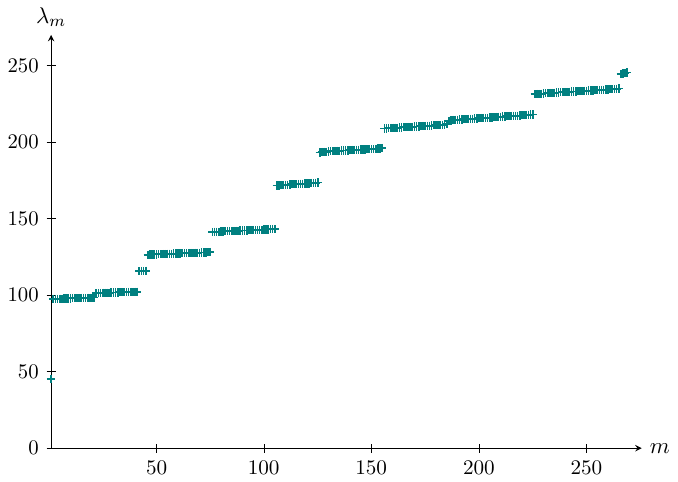}
\caption{A plot of the numerical eigenvalues of the Laplacian on the Fermat quintic for $(3,0)$-forms with $k_{\phi}=3$ and $N_{\phi}=3\times{10}^{6}$.}
\end{figure}

\begin{figure}
\includegraphics{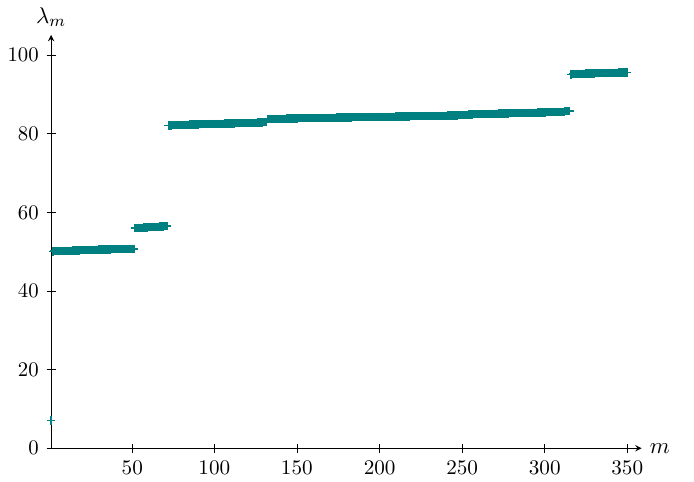}
\caption{A plot of the numerical eigenvalues of the Laplacian on the Fermat quintic for $(1,1)$-forms with $k_{\phi}=3$ and $N_{\phi}=3\times{10}^{6}$.}
\end{figure}

\begin{figure}
\includegraphics{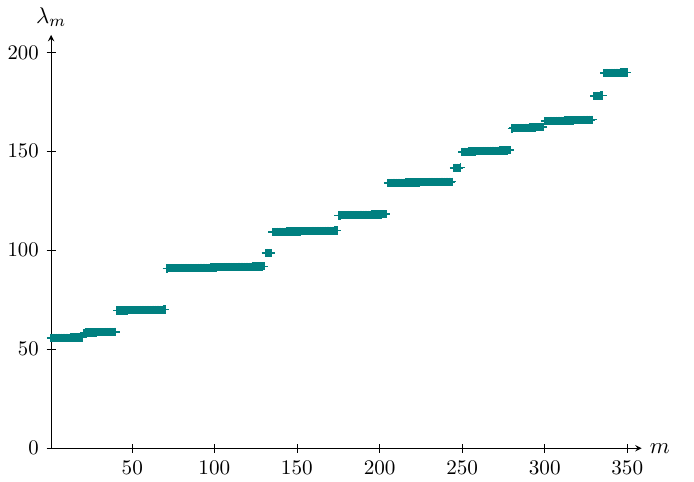}
\caption{A plot of the numerical eigenvalues of the Laplacian on the Fermat quintic for $(2,1)$-forms with $k_{\phi}=3$ and $N_{\phi}=3\times{10}^{6}$.\label{fig:extra2}}
\end{figure}

\bibliographystyle{utphys}
\bibliography{main}

\end{document}